\documentclass[%
preprint,
 amsmath,amssymb,
 aps,
]{revtex4-2}

\usepackage{graphicx}
\usepackage{dcolumn}
\usepackage{bm}
\renewcommand\bf\bfseries  

\begin{document}

\title{Revised Hamiltonian near Third-Integer Resonance and Implications for Transverse Resonance Island Buckets }
\author{Ki Moon Nam}
\affiliation{Division of Advanced Nuclear Engineering, Pohang University of Science and Technology, Pohang 37673, Rep. of Korea}
\author{Jaehyun Kim}
\affiliation{Pohang Accelerator Laboratory, Pohang University of Science and Technology, Pohang 37673, Rep. of  Korea}
\author{Young Dae Yoon}
\email{youngdae.yoon@apctp.org}
\affiliation{Asia Pacific Center for Theoretical Physics, Pohang 37673, Rep. of Korea}
\author{Yong Woon Parc}
\email{young1@postech.ac.kr}
\affiliation{Pohang Accelerator Laboratory, Pohang University of Science and Technology, Pohang 37673, Rep. of  Korea}
\affiliation{Division of Advanced Nuclear Engineering, Pohang University of Science and Technology, Pohang 37673, Rep. of Korea}

 





\date{\today}

\begin{abstract}
In storage rings, an accurate description of particle dynamics near third-integer resonance is crucial for various applications. The conventional approach is to extrapolate far-resonance dynamics to near-resonance, but difficulty arises because the nonlinear detuning parameter diverges at this critical point. Here we derive, via a suitable application of the canonical perturbation theory, a revised detuning parameter that is well-behaved near resonance. The resultant theory accurately describes the morphology of transverse resonance island buckets (TRIBs) for a wide range of parameter space. Our results have important implications for advanced applications of storage rings, as well as for the underlying physics of resonant particle dynamics.
\end{abstract}

\maketitle


\section{INTRODUCTION}
A charged particle in a storage ring experiences a periodic potential which makes the particle susceptible to various resonances.
Such resonance phenomena have traditionally been viewed as detrimental to beam stability, and operating the storage ring near resonance tunes has been avoided.
However, various means are being revisited to actually utilize this resonance phenomenon to applications such as slow extraction \cite{1Nagaslaev} and autoresonant excitation \cite{2Minghao}.

One such application is the Transverse Resonance Island Buckets (TRIBs) where the charged particles are confined to specific regions in phase space by generating or amplifying certain resonance \cite{3Paul}.
Sextupole magnets in storage rings can form additional island buckets surrounding the central primary bucket.
Previous research focus had been on eliminating these buckets \cite{26robin}, but particles can actually be trapped into the island buckets by choosing the appropriate tune and applying external kicks \cite{4Kim}, allowing for a multi-objective utilization of the stored beam.
TRIBs has been implemented at several facilities such as BESSY II \cite{3Paul} and MAX IV \cite{5Olsson}.
The presence of multiple stable orbits in the ring has enabled pump-and-probe experiments with spatially separated short X-ray pulses  \cite{6Hwang}, synchrotron-radiation-based electron time-of-flight spectroscopy \cite{7Arion}, and control of X-ray helicity using APPLE-type undulators in conjunction with TRIBs operation \cite{8Holldack}. 

Despite the experimental implementation of TRIBs, the theoretical description of this phenomenon is still a subject of ongoing research and not fully understood \cite{9Marc}.
A widely-used dynamical framework in storage ring physics is the Hamiltonian formalism \cite{10Mello,11Guignard,12Wiedemann}.
When investigating higher-order effects beyond linear storage ring dynamics, a common approach is to separate long term and short term motions to derive an effective or average Hamiltonian that describes the system's long-term behavior and driving mechanisms.
This mathematical approach has been applied to study amplitude dependent tune shifts \cite{13Soutome} and nonlinear chromaticity \cite{14Takao} in storage rings. 

A key aspect for understanding TRIBs is the dynamical properties near the tune \(\nu_{x} \cong l_{3\nu_{x}}/3\), which corresponds to the third-integer resonance.
Near this resonance, the Hamiltonian has been proposed as \cite{15Lee}
\begin{equation}
\label{doc1}
\mathcal{H}(\phi,J) = \left( \nu_{x} - \frac{l_{3{\nu}_{x}}}{3} \right)J + g_{3,0,l_{3\nu_x}}J^{\frac{3}{2}}\cos(3\phi) + \frac{1}{2}\alpha_{0}J^{2},\ 
\end{equation}
where \((\phi,\ J)\) are the action-angle variables, \(\nu_{x}\) is the horizontal tune, \(l_{3\nu_{x}}\) is the integer number closest to \(3\nu_{x}\), \(g_{3,0,l_{3\nu_x}}\) is the resonance strength, and \(\alpha_{0}\) is the nonlinear detuning parameter \cite{15Lee,16Merminga}. 
However, \(\alpha_{0}\) diverges as \(\nu_{x}\rightarrow \frac{l_{3\nu_{x}}}{3}\), and so this theory breaks down near the third-integer resonance around which TRIBs mode is supposed to operate.

In this Letter, we present a revised expression for the nonlinear detuning parameter using perturbative canonical transformations.
The revised parameter is well-behaved near third-integer resonance and so accurately describes the presence and morphology of the additional islands in TRIBs mode.
Particle tracking simulations using the lattice information of a currently-operating storage ring (PLS-II) are performed, and their results are shown to conform to the analytical predictions.
The bearing of our findings on advanced operations of storage rings is discussed.

\section{HAMILTONIAN FOR A STORAGE RING WITH SEXTUPOLE MAGNET}
The coordinate system (Frenet-Serret) employed in this study is depicted in Fig. \ref{figa1}.
The Hamiltonian, as given in Eq. (6) of Ref. \cite{13Soutome}, is presented below:
\begin{align}
\displaystyle \mathcal{H}_1 &= \displaystyle \frac{I_x}{\beta_x(s)}+V\left(\phi_x,I_x,s\right), \\
\displaystyle  V\left(\phi_x,I_x,s\right)&= \displaystyle  \frac{m_x(s)}
{6\sqrt2}\left(\sqrt{\beta_xI_x}\right)^3\left(\cos\left(3\phi_x\right)+3\cos(\phi_x)\right),   
\end{align}
where \(\beta_{x}(s)\) is a horizontal betatron function, \( \left(\phi_x,I_x\right) \) 
 is action-angle variables and the sextupole magnet strength \(m_x(s)\) is given by
\begin{equation}
m_x(s)=\frac{e}{p}\frac{\partial^2B}{\partial x^2}.
\end{equation}
In the above definitions, \(p\) is the momentum of an electron, \(e\) is the charge of an electron, \(\rho\) is the bending radius and \(B\) is the magnetic field strength of the sextupole magnet.

A canonical transformation mapping from \((\phi_x,I_x)\) to \((\psi_2,J_2)\) is performed using the second type of generating function, as given by \cite{15Lee}
\begin{equation}
F_2\left(\phi_x,J_2,s\right)=\left(\phi_x-\int_{0}^{s}{\frac{1}{\beta_x\left(\tau\right)}d\tau}+\frac{2\pi}{L}s\nu_x\right)J_2,   
\end{equation}
where \(L\) is the periodicity in the storage ring (for example, the length of a lattice or the circumference) and \(\nu_x\) is defined as
\begin{equation}
\nu_x=\frac{1}{2\pi}\int_{0}^{L}\frac{1}{\beta_x\left(\tau\right)}d\tau.
\end{equation}
The new canonical action-angle variables are given below:
\begin{align}
\displaystyle I_x & \displaystyle=\frac{\partial F_2}{\partial\phi_x}=J_2, \\
\displaystyle \psi_2 & =\frac{\partial F_2}{\partial J_2}=\phi_x-\int_{0}^{s}{\frac{1}{\beta_x\left(\tau\right)}d\tau}+\frac{2\pi}{L}s\nu_x,
\end{align}
where the numerical subscript signifies the number of canonical transformations from the \(\left(x,x^\prime\right)\) position-momentum space. 
From the definition of generating function and replacing the system variable from \(s\) to \(\displaystyle \theta=\frac{2\pi s}{L}=\frac{s}{R}\), then the transformed Hamiltonian is given by 
\begin{equation}
\mathcal{H}_2\left(\psi_2,J_2,\theta\right)=\nu_xJ_2+V\left(\psi_2,J_2,\theta\right),   
\end{equation}
where
\begin{equation}
\begin{array}{cl}
\displaystyle V\left(\psi_2,J_2,\theta\right) = & \displaystyle \frac{Rm_x(\theta)}{6\sqrt2}\left[\left(\sqrt{\beta_xJ_2}\right)^3\cos\left(3\psi_2-3\nu_x\theta+3\chi_x(\theta)\right)\right.\\
\displaystyle &\displaystyle  \left.+3\left(\sqrt{\beta_xJ_2}\right)^3\cos\left(\psi_2-\nu_x\theta+\chi_x(\theta)\right)\right],
\end{array}    
\end{equation}
and
\begin{equation}
\chi_x(\theta)\equiv R\int_{0}^{\theta}{\frac{1}{\beta_x\left(R\tau\right)}d\tau}.
\end{equation}
\begin{figure}
\includegraphics[width=7cm]{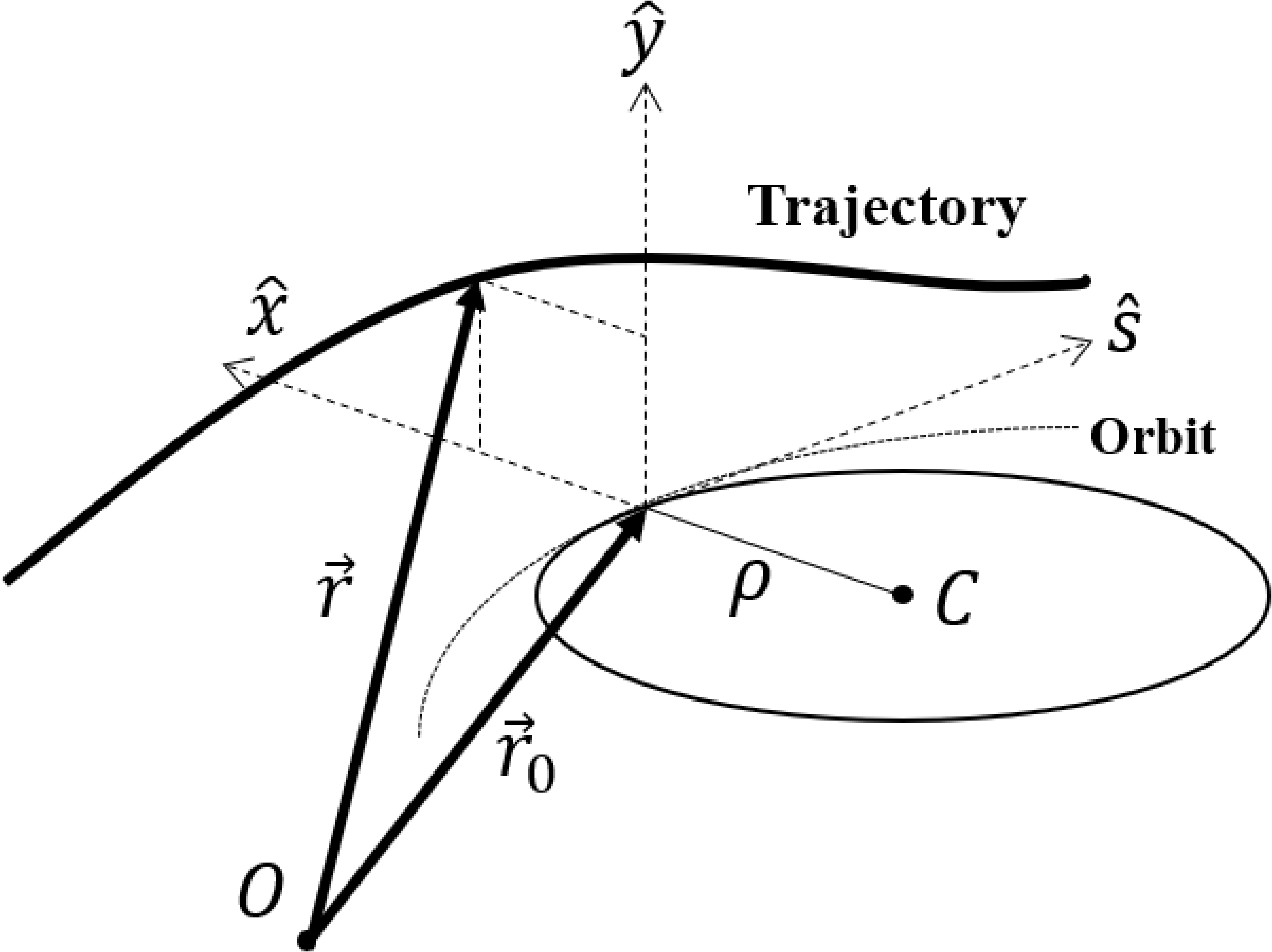}
\caption{\label{figa1} 
The Frenet-Serret coordinate system used in this study.
The particle moves along the trajectory line, with the position denoted by the vector \(\vec{r}\).
The origin of the Frenet-Serret coordinate system is also denoted by the vector \(\protect\overrightarrow{r_0}\).
The ideal orbit is represented by dotted curved line.
The bending radius is denoted by \(\rho\), and the unit vectors of each axis are denoted by \(\hat{x}\), \(\hat{y}\), and \(\hat{s}\).}
\end{figure}
Then, Fourier expanding \(V\) in \(\theta\), the Hamiltonian is now
\begin{equation}
\begin{array}{cl}
     \mathcal{H}_2\left(\psi_2,J_2,\theta\right) = & \nu_xJ_2+\left(\sqrt{J_2}\right)^3\displaystyle\sum_{n=-\infty}^{\infty}{g_{3,0,n}\cos{\left(3\psi_2-n\theta+\xi_{3,0,n}\right)}} \\
     & +\left(\sqrt{J_2}\right)^3\displaystyle\sum_{n=-\infty}^{\infty}{g_{1,0,n}\cos{\left(\psi_2-n\theta+\xi_{1,0,n}\right)}},
\end{array}    
\end{equation}
where the Fourier coefficients \(g_{3,0,n}\), \(\xi_{3,0,n}\), \(g_{1,0,n}\), \(\xi_{1,0,n}\) are given in the APPENDIX A.

We now perform a canonical transformation using the generating function
\begin{equation}
G\left(\psi_2,J_3,\theta\right)=\left(\psi_2-\frac{l_{3\nu_x}}{3}\theta\right)J_3,
\end{equation}
which in effect eliminates the linear $\theta$-dependency of the angle variable. Then, new Hamiltonian is given as
\begin{equation}
    \mathcal{H}_3\left(\psi_3,J_3,\theta\right)=\mathcal{H}_2\left(\psi_3,J_3,\theta\right)+\frac{\partial G\left(\psi_2,J_3,\theta\right)}{\partial\theta}=\delta_\nu J_3+V\left(\psi_3,J_3,\theta\right),
\end{equation}
where the resonance proximity parameter \(\delta_\nu\) is defined by
\begin{equation}
\delta_\nu\equiv\nu_x-\frac{l_{3\nu_x}}{3}    
\end{equation}
and potential term is given by

\begin{equation}
\begin{array}{cl}
\label{doc5}   
V\left(\psi_3,J_3,\theta\right) = & \displaystyle \left(\sqrt{J_3}\right)^3\displaystyle\sum_{n=-\infty}^{\infty}{g_{3,0,n}\cos{\left(3\psi_3+\left(l_{3\nu_x}-n\right)\theta+\xi_{3,0,n}\right)}}\\
     & \displaystyle +\left(\sqrt{J_3}\right)^3\displaystyle\sum_{n=-\infty}^{\infty}{g_{1,0,n}\cos{\left(\psi_3+\left(\frac{l_{3\nu_x}}{3}-n\right)\theta+\xi_{1,0,n}\right)}}.
\end{array}    
\end{equation}
Here, \(\frac{d\psi_3}{d\theta}=\frac{\partial\mathcal{H}_3}{\partial J_3}=\delta_\nu+\frac{\partial V}{\partial J_3}\) and \(\delta_\nu\ll1\) near resonance, so if \(V\) is assumed to be of first-order in smallness, \(\psi_3\) is a slowly varying function of \(\theta\). 
 Then, Eq. (\ref{doc5}) shows that \(V\) consists of fast-varying terms that depend on \(\theta\) and a slowly varying term for \(n=l_{3\nu_x}\) that does not depend on \(\theta\).
\section{CANONICAL PERTURBATION AND \(\theta\)-INDEPENDENT HAMILTONIAN \(\mathcal{H}_4\)}

Now we perform another perturbative canonical transformation from \(\left(\psi_3,J_3\right)\) to 
  \(\left(\psi,J\right)\) that renders the transformed Hamiltonian to be explicitly \(\theta\)-invariant for up to second order in smallness \cite{13Soutome,18Ruth}.
The generating function can be written as 
\begin{equation}    F\left(\psi_3,J,\theta\right)=\psi_3J+F^{(1)}\left(\psi_3,J,\theta\right)+F^{(2)}\left(\psi_3,J,\theta\right)+\ldots,
\end{equation}
where the superscript signifies the order of the perturbation.
The transformed action variable \(J\)  is now determined by the following relation: 
\begin{equation}
J_3=\frac{\partial F\left(\psi_3,J,\theta\right)}{\partial\psi_3}=J+\frac{\partial F^{(1)}\left(\psi_3,J,\theta\right)}{\partial\psi_3}+\frac{\partial F^{(2)}\left(\psi_3,J,\theta\right)}{\partial\psi_3}\ .    
\end{equation}
The Hamiltonian \(\mathcal{H}_4\) is given by:
\begin{equation}
\label{H4}
\displaystyle \mathcal{H}_4=\mathcal{H}_3+\frac{\partial F(\psi_3,J,\theta)}{\partial \theta},
\end{equation}
By using a Taylor series, the Hamiltonian $\mathcal{H}_4$ can be arranged in order of their smallness and is given up to second order by,
\begin{equation}
\mathcal{H}_4\left(\psi_3,J,\theta\right)\cong\mathcal{H}^{\left(0\right)}+\mathcal{H}^{\left(1\right)}+\mathcal{H}^{\left(2\right)},
\end{equation}
where
\begin{align}
\mathcal{H}^{\left(0\right)}&\equiv\delta_\nu J,  \label{doc7}\\
\mathcal{H}^{(1)}&\equiv V(\psi_3,J,\theta)+\delta_\nu\frac{\partial F^{\left(1\right)}}{\partial\psi_3}+\frac{\partial F^{\left(1\right)}}{\partial\theta},\label{doc8} \\
\mathcal{H}^{\left(2\right)}&\equiv\frac{\partial V(\psi_3,J,\theta)}{\partial J}\frac{\partial F^{\left(1\right)}}{\partial\psi_3}+\delta_\nu\frac{\partial F^{\left(2\right)}}{\partial\psi_3}+\frac{\partial F^{\left(2\right)}}{\partial\theta}. \label{doc9}
\end{align}
Here we have used the fact that the sextupole strength is small enough so that \(V\) is of first order.
The \(\theta\)-invariance of \(\mathcal{H}^{(n)}\) for \(n\in \{0,1,2\}\) means that the \(n\)-th order generating function \(F^{(n)}\) should satisfy 
\begin{equation}
\label{doc10}
\mathcal{H}^{(n)}-\left<{\mathcal{H}^{(n)}}\right>_{\theta}=0,   
\end{equation}
where \(\left<{A}\right>_{\theta}\) means the average of \(A\) over \(\theta\). 
Note that \(\mathcal{H}^{(0)}\) is already \(\theta\)-invariant, so we start from \(n=1\).

For the first-order Hamiltonian \(\mathcal{H}^{(1)}\), it should satisfy
\begin{equation}
\label{H1c}
\mathcal{H}^{(1)}-\left<{\mathcal{H}^{(1)}}\right>_{\theta}=0.
\end{equation}
The \(\theta\)-average of the first-order Hamiltonian is given by
\begin{equation}
\begin{array}{cl}
\left<{\mathcal{H}^{\left(1\right)}}\right>_{\theta}&=\displaystyle\left<{V\left(\psi_3,J,\theta\right)+\delta_\nu\frac{\partial F^{(1)}}{\partial\psi_3}+\frac{\partial F^{(1)}}{\partial\theta}}\right>_{\theta}\\
     &=\displaystyle\left<{V\left(\psi_3,J,\theta\right)}\right>_{\theta}+\delta_\nu\left<{\frac{\partial F^{(1)}}{\partial\psi_3}}\right>_{\theta}+\left<{\frac{\partial F^{(1)}}{\partial\theta}}\right>_{\theta}.
     
\end{array}    
\end{equation}
All terms in \(\left<{V\left(\psi_3,J,\theta\right)}\right>_{\theta}\) are \(0\) except one term because other terms in \(V\left(\psi_3,J,\theta\right)\) has explicit oscillatory dependency on \(\theta\). 
In order to satisfy \(\mathcal{H}^{(1)}=\left<{\mathcal{H}^{(1)}}\right>_{\theta}\), we need to find the generating function that satisfies the following relations:
\begin{eqnarray}
\left<{\frac{\partial F^{(1)}}{\partial\psi_3}}\right>_{\theta}=0,\\
\left<{\frac{\partial F^{\left(1\right)}}{\partial\theta}}\right>_{\theta}=0.
\end{eqnarray}
Assuming that the above two equations are satisfied by some generating function, the first-order Hamiltonian is given by
\begin{equation}
\label{H1ave}
\begin{array}{cl}
\mathcal{H}^{(1)}&=\left<{\mathcal{H}^{(1)}}\right>_{\theta} \\
&=
\displaystyle\left<{V\left(\psi_3,J,\theta\right)}\right>_{\theta}+\delta_\nu\left<{\frac{\partial F^{(1)}}{\partial\psi_3}}\right>_{\theta}+\left<{\frac{\partial F^{(1)}}{\partial\theta}}\right>_{\theta}\\
&=
\left(\sqrt J\right)^3g_{3,0,l_{3\nu_x}}\cos\left(3\psi_3+\xi_{3,0,l_{3\nu_x}}\right).
\end{array} 
\end{equation}
From Eqs. (\ref{doc8}) and (\ref{H1ave}), we can derive the following equation:
\begin{equation}
\label{1ordergendifeqn}
\left\{\delta_\nu\frac{\partial}{\partial\psi_3}+\frac{\partial}{\partial\theta}\right\}F^{(1)}=
-\left(V\left(\psi_3,J,\theta\right)-\left(\sqrt J\right)^3g_{3,0,l_{3\nu_x}}\cos{\left(3\psi_3+\xi_{3,0,l_{3\nu_x}}\right)}\right).
\end{equation}
Using above equation, we can determine the first-order generating function, \(F^{(1)}\left(\psi_3,J,\theta\right)\). We try the following ansatz for the generating function based on the form of \(V\left(\psi_3,J,\theta\right)\):
\begin{equation}
\begin{array}{cl}
\label{F1}
\displaystyle F^{\left(1\right)}\left(\psi_3,J,\theta\right) = & \displaystyle J^\frac{3}{2}\sum_{n=-\infty}^{\infty}{\biggl( f_{3,0,n}\sin\left(3\psi_3-\left(n-l_{3\nu_x}\right)\theta+\xi_{3,0,n}\right) }   \\
&\displaystyle +f_{1,0,n}\sin\left(\psi_3-\left(n-\frac{l_{3\nu_x}}{3}\right)\theta+\xi_{1,0,n}\right) \biggr).\\
\end{array}
\end{equation}
We can obtain the following relations from Eqs. (\ref{1ordergendifeqn}-\ref{F1}):
\begin{equation}
3\delta_\nu f_{3,0,n}+f_{3,0,n}\left(l_{3\nu_x}-n\right)=-g_{3,0,n},
\end{equation}
and
\begin{equation}
\delta_\nu f_{1,0,n}+f_{1,0,n}\left(\frac{l_{3\nu_x}}{3}-n\right)=-g_{1,0,n}.
\end{equation}
Thus, using the definition of \(\delta_{\nu}\), we obtain the coefficients of the first-order generating function as follows:
\begin{equation}
\label{olc16}
f_{a,0,n}=-\frac{g_{a,0,n}}{a\nu_x-n}\ \ \ \mathrm{for} \ a\in\left\{1,3\right\}.   
\end{equation}

Now, we can calculate the second-order Hamiltonian using the following relation: 
\begin{equation}
\mathcal{H}^{(2)}-\left<{\mathcal{H}^{(2)}}\right>_{\theta}=0.
\end{equation}
The average value of the second-order Hamiltonian is calculated as follows:
\begin{equation}
\left<{\mathcal{H}^{\left(2\right)}}\right>_{\theta}
=\delta_\nu\left<{\frac{\partial F^{(2)}}{\partial\psi_3}}\right>_{\theta}
+\left<{\frac{\partial F^{(2)}}{\partial\theta}}\right>_{\theta}
+\left<{\frac{\partial V\left(\psi_3,J,\theta\right)}{\partial J}\frac{\partial F^{(1)}}{\partial\psi_3}}\right>_{\theta}.
\end{equation}
Following the same way as in the first-order Hamiltonian, we can assume the following:
\begin{eqnarray}
\left<{\frac{\partial F^{(2)}}{\partial\psi_3}}\right>_{\theta}=0,\\
\left<{\frac{\partial F^{(2)}}{\partial\theta}}\right>_{\theta}=0.
\end{eqnarray}
In this step, it is not necessary to explicitly calculate \(F^{(2)}\left(\psi_3,J,\theta\right)\), and it is enough to assume that the \(\theta\)-average of the second-order generating function is zero for the purposes of this study.
Consequently, the second-order Hamiltonian contains only \(\theta\)-invariant terms, which can be calculated to obtain the resulting expression:
\begin{equation}
\begin{array}{cl}
\displaystyle\left<{\mathcal{H}^{\left(2\right)}}\right>_{\theta} & =  \displaystyle \left<{\frac{\partial V\left(\psi_3,J,\theta\right)}{\partial J}\frac{\partial F^{(1)}}{\partial\psi_3}}\right>_{\theta} \\
&\displaystyle =  \frac{3}{4}J^{2}\left<\sum_{n,n^\prime=-\infty,n^\prime\neq l_{3\nu_x}}^{\infty}{3g_{3,0,n}f_{3,0,n^\prime}\cos\left(-\left(n-n^\prime\right)\theta+\xi_{3,0,n}-\xi_{3,0,n^\prime}\right)}\right. \\
&\displaystyle~+\sum_{n,n^\prime=-\infty}^{\infty}{g_{1,0,n}f_{1,0,n^\prime}\cos\left(-\left(n-n^\prime\right)\theta+\xi_{1,0,n}-\xi_{1,0,n^\prime}\right)}\\
& \displaystyle ~\left.+\sum_{n,n^\prime=-\infty}^{\infty}{f_{3,0,n}g_{3,0,n^\prime}\cos{\left(6\psi_3-\left(n+n^\prime-2l_{3\nu_x}\right)\theta\right)}\cos{\xi_n}\cos{\xi_n^\prime}}\right>_{\theta}\\
\displaystyle &=\displaystyle \frac{3}{4}{J}^{2} \left\{\sum_{n=-\infty,n\neq l_{3\nu_{x}}}^{\infty}{3f_{3,0,n}g_{3,0,n}}+\sum_{n=-\infty}^{\infty}{f_{1,0,n}g_{1,0,n}} \right.\\
&\displaystyle \left.+\cos{\left(6\psi_3\right)}\sum_{n+n^\prime=2l_{3\nu_x}}{f_{3,0,n}g_{3,0,n^\prime}\cos{\xi_n}\cos{\xi_{n^\prime}}} \right\}.
\end{array}    
\end{equation}
The coefficient of \(\cos{\left(6\psi_3\right)}\) can be expressed in the following form:
\begin{equation}
\label{coef6psi}
\begin{array}{cl}
\displaystyle\sum_{n+n^\prime=2l_{3\nu_x}}{f_{3,0,n}g_{3,0,n^\prime}\cos{\xi_n}\cos{\xi_{n^\prime}}}
&=\displaystyle\sum_{k=-\infty}^{\infty}{f_{3,0,l_{3\nu_x}-k}g_{3,0,l_{3\nu_x}+k}\cos{\xi_{l_{3\nu_x}-k}}\cos{\xi_{l_{3\nu_x}+k}}}\\
&=\displaystyle\sum_{k=-\infty,k\neq 0}^{\infty}{\frac{g_{3,0,l_{3\nu_x}+k}g_{3,0,l_{3\nu_x}-k}}{l_{3\nu_x}+k-3\nu_x}\cos{\xi_{l_{3\nu_x}+k}}\cos{\xi_{l_{3\nu_x}-k}}}\\
&\cong\displaystyle\sum_{k=-\infty,k\neq 0}^{\infty}{\frac{g_{3,0,l_{3\nu_x}+k}g_{3,0,l_{3\nu_x}-k}}{k}\cos{\xi_{l_{3\nu_x}+k}}\cos{\xi_{l_{3\nu_x}-k}}},
\end{array}    
\end{equation}
where we used \(l_{3\nu_x}\cong3\nu_x\). 
The function under the summation in Eq. (\ref{coef6psi}) is odd with respect to $k$, so the value of the sum is zero.
Therefore, the second-order Hamiltonian can be obtained as follows:
\begin{equation}
\label{H2ave}
\begin{array}{cl}
\displaystyle \mathcal{H}^{(2)}
& \displaystyle  =\left<{\mathcal{H}^{(2)}}\right>_{\theta}\\
& \displaystyle =
\left<{\frac{\partial V\left(\psi_3,J,\theta\right)}{\partial J}\frac{\partial F^{(1)}}{\partial\psi_3}}\right>_{\theta} \\
& \displaystyle =\frac{3}{4}J^{2}\left(\sum_{n=-\infty,n\neq l_{3\nu_{x}}}^{\infty}{3f_{3,0,n}g_{3,0,n}}+\sum_{n=-\infty}^{\infty}{f_{1,0,n}g_{1,0,n}}\right) \\
& \displaystyle=\frac{1}{2}\alpha_{-1}J^2.
\end{array} 
\end{equation}
The full Hamiltonian can be obtained from Eqs. (\ref{doc7}), (\ref{H1ave}) and (\ref{H2ave}) as follows:
\begin{equation}
\displaystyle \mathcal{H}_4\left(\psi_3,J\right)\cong\delta_\nu J+J^\frac{3}{2}g_{3,0,l_{3\nu_x}}\cos\left(3\psi_3+\xi_{3,0,l_{3\nu_x}}\right)+\frac{1}{2}\alpha_{-1}J^2.
\end{equation}
To express the Hamiltonian in terms of new variables, we use the following relation between the old and new angle variables:
\begin{equation}
\psi\cong\psi_3+\frac{\partial F^{\left(1\right)}\left(\psi_3,J,\theta\right)}{\partial J},
\end{equation}
where the last term is of first order in smallness, i.e., $\Delta\psi$. However,
\begin{equation}
\begin{array}{cl}
\displaystyle\cos\left(3\psi_3+\xi_{3,0,l_{3\nu_x}}\right)&=\displaystyle\cos\left(3\psi+\xi_{3,0,l_{3\nu_x}}-\Delta\psi\right)\\
&\displaystyle=\cos\left(3\psi+\xi_{3,0,l_{3\nu_x}}\right)+O\left(\Delta\psi^2\right)\\
&\displaystyle\cong \cos\left(3\psi+\xi_{3,0,l_{3\nu_x}}\right).
\end{array}    
\end{equation}
Therefore, the resulting \(\theta\)-invariant Hamiltonian \(\mathcal{H}_4\) is obtained as follows:
\begin{equation}
\label{doc11}
\mathcal{H}_4\left(\psi,J\right)=\delta_\nu J+g_{3,0,l_{3\nu_x}}J^\frac{3}{2}\cos\left(3\psi+\xi_{3,0,l_{3\nu_x}}\right)+\frac{1}{2}\alpha_{-1}J^2,
\end{equation}
where
\begin{equation}
\label{doc12}
\displaystyle  \alpha_{-1}=\frac{3}{2}\left(\sum_{n=-\infty,n\neq l_{3\nu_x}}^{\infty}{3f_{3,0,n}g_{3,0,n}}+\sum_{n=-\infty}^{\infty}{f_{1,0,n}g_{1,0,n}}\right).
\end{equation}
It should be emphasized that the omitted term in the summation in Eq. (\ref{doc12}) corresponds to the slowly varying term in the first-order Hamiltonian in Eq. (\ref{doc8}).
By comparison to Eq. (\ref{doc1}), \(\alpha_{-1}\) is the revised detuning parameter and is one of our main results.
The first (unperturbed), second (resonance-driving), and third (detuning and island-forming) terms in Eq. (\ref{doc11}) correspond to \(\mathcal{H}^{\left(0\right)}\), \(\mathcal{H}^{\left(1\right)}\), and \(\mathcal{H}^{\left(2\right)}\), respectively.

In contrast, the conventional detuning parameter in Eq. (1) is effectively given by,
\begin{equation}
\label{doc13}
\displaystyle  \alpha_0=\frac{3}{2}\left(\sum_{n=-\infty}^{\infty}{3f_{3,0,n}g_{3,0,n}}+\sum_{n=-\infty}^{\infty}{f_{1,0,n}g_{1,0,n}}\right).   
\end{equation}
The reason why Eq. (\ref{doc13}) corresponds to Eq. (63) in Ref. \cite{13Soutome} or Eq. (196) in Ref. \cite{15Lee} is given in Appendix B. Equation (\ref{doc13}) is derived by averaging over both the faster-varying \(\theta\) and the slowly varying \(\psi_3\) in Eq. (\ref{doc10}). 
The relation between Eq. (\ref{doc12}) and (\ref{doc13}) is given by 
\begin{equation}
\label{doc14}
\displaystyle \alpha_0=\alpha_{-1}+\frac{9}{2}\frac{{g_{3,0,l_{3\nu_x}}}^2}{l_{3\nu_x}-3\nu_x}\ .   
\end{equation}
It is clear that the last term of Eq. (\ref{doc14}) which is the omitted term in Eq. (\ref{doc12}) diverges when the tune \(\nu_x\) is close to \(l_{3\nu_x}/3\). Because of this omission, \(\alpha_{-1}\) is well-behaved and correctly describes near-resonance dynamics.

The analytical prediction given by \(\alpha_{-1}\) will now be verified through comparisons to numerical simulations.
An electron tracking code was written in MATLAB that treats dipole and quadrupole magnets as transfer matrices and solves sextupole effects using the fourth-order Runge-Kutta method. 
The algorithm was tested against the PLS-II storage ring lattice, shown in Fig. \ref{fig1} \cite{21PLSII}.
To facilitate the comparison, we define the following quantities:
\begin{align}
\displaystyle X&=\sqrt J\cos{\left(\psi\right)},\\
\displaystyle P&=-\sqrt J\sin{\left(\psi\right)}.
\end{align}
Equation (\ref{doc11}) is now given by, 
\begin{equation}
\label{doc15}
\displaystyle \mathcal{H}_5\left(X,P\right)=\delta_\nu\left\{X^2+P^2\right\}+g_{3,0,l_{3\nu_x}}\left\{X^3-3XP^2\right\}\cos{\left(\xi_{3,0,l_{3\nu_x}}\right)}+\frac{1}{2}\alpha_j\left\{X^2+P^2\right\}^2,   
\end{equation}
where \(j=-1\) yields the revised detuning parameter in Eq. (\ref{doc12}), and \(j=0\) yields the conventional detuning parameter in Eq. (\ref{doc13}).

\section{NUMERICAL RESULTS}
\begin{figure}
\includegraphics[width=7cm]{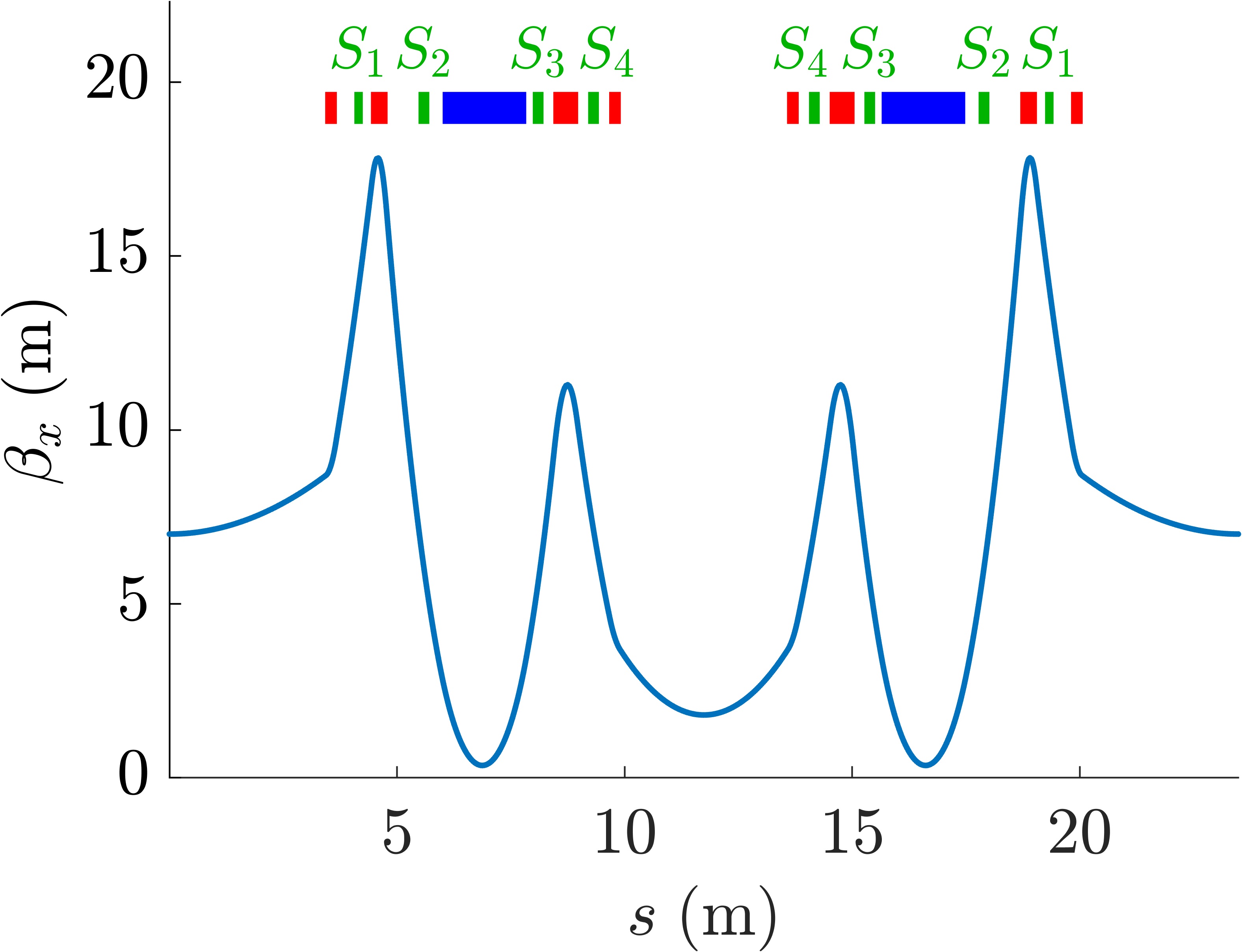}
\caption{\label{fig1}  PLS-II lattice and horizontal beta function. PLS-II is a ‘Double Bend Achromats (DBA)’ lattice. The initial horizontal beta function is \(\beta_{x,0} = 7.006\) \(\mathrm{m}\) and the tune is \(\nu_x = 1.3325\).
The rectangular inset illustrates the magnet distribution within PLS-II, where the red denotes the quadrupole magnets, the blue the bending magnets, and the green the sextupole magnets \(S_{1-4}\).}
\end{figure}

\begin{figure}
\includegraphics[width=7cm]{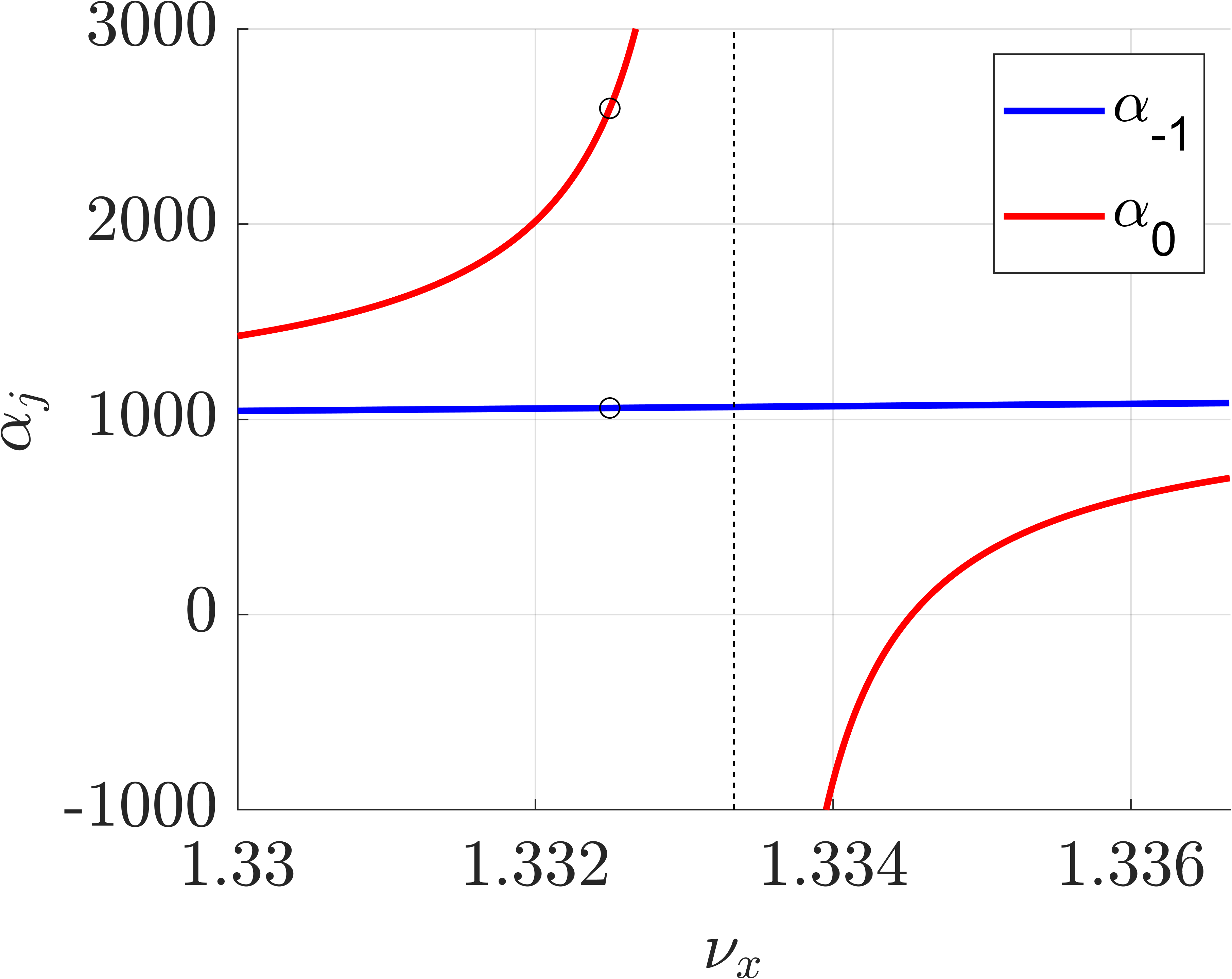}
\caption{\label{fig2}  Nonlinear detuning parameter vs tune. 
The original detuning parameter \(\alpha_0\) is plotted by red line.
The revised detuning parameter \(\alpha_{-1}\) is plotted by blue line. The circles denote the values of detuning parameters when the tune is \(\nu_{x}=1.3325\) for PLS-II lattice.}
\end{figure}

While the designed tune of the PLS-II lattice is 1.273, the fiducial tune was set to 1.3325 to form TRIBs. There are four pairs of sextupole magnets (green boxes in Fig. \ref{fig1}) whose strengths determine the values of \(\alpha_j\).
The behavior of \(\alpha_j\) for the PLS-II lattice as a function of \(\nu_x\) around the fiducial \(\nu_x\) is presented in Fig. \ref{fig2}.
The original detuning parameter \(\alpha_0\) (red line) diverges when the fractional tune is close to \(1/3\) (vertical dashed line) while the revised parameter \(\alpha_{-1}\) (blue line) is well-behaved near \(1/3\). At the fiducial tune \(\nu_x = 1.3325\), \(\alpha_0\) is \(2594.2\) and \(\alpha_{-1}\) is \(1058.1\) (circles in the Fig. \ref{fig2}).

\begin{figure}
\includegraphics[width=7cm]{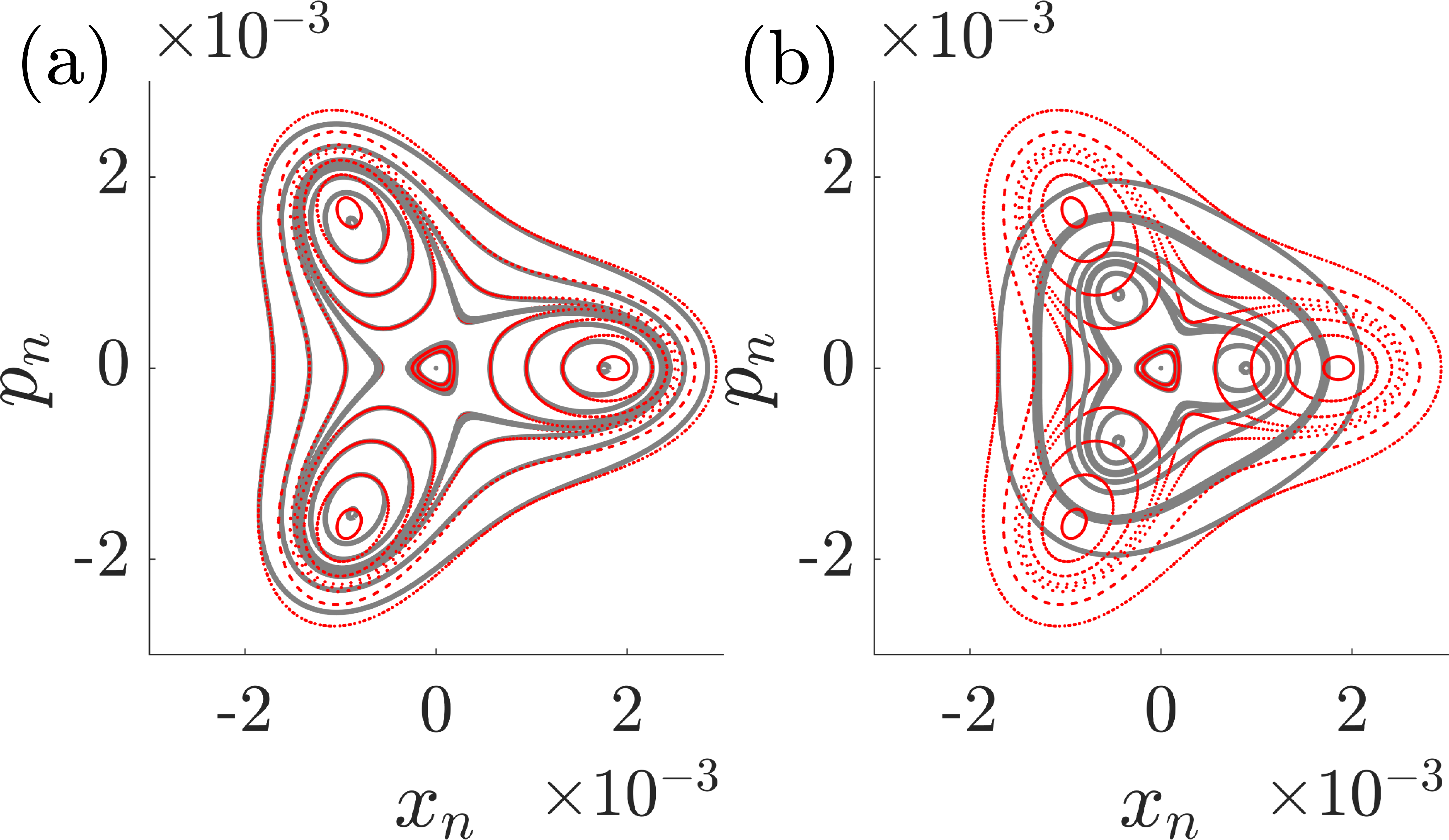}
\caption{\label{fig3}  (a) Phase space trajectories with contour plots of the Hamiltonian in Eq. (\ref{doc15}) where \(\delta=-8.33 \ \times \ 10^{-4}\), \(g_{3,0,l_{3\nu_x}}=-0.923\) and \(j=-1\) (gray line). 
Red dots represent phase space trajectories from tracking results.
Normalized coordinates are defined as \(\left(x_n=\frac{x}{\sqrt{2\beta_{x,0}}},p_n=-\sqrt{\frac{\beta_{x,0}}{2}}x^\prime\right)\) where \(\beta_{x,0}\) is the beta function at the starting position of the lattice. (b) Same as (a), but for \(j=0\). }
\end{figure}

The simulated electron phase-space trajectories are shown as a Poincare section (red) in Fig. \ref{fig3}. 
10 electrons were initiated with \(\left(X,P\right)=\left(X_0,0\right)\), where \(X_0\) uniformly ranges from -0.0017 to 0.0017 in normalized phase space.
As they pass through the periodic lattice, their phase-space positions at the start of the lattice were recorded for 1500 cells. 
Also plotted in gray is a contour plot of Hamiltonian in Eq. (\ref{doc15}) with \(\alpha_{-1}\) (Fig. \ref{fig3}(a)) and \(\alpha_{0}\) (Fig. \ref{fig3}(b)).
There is good agreement between the Poincare section and the gray contour using $\alpha_{-1}$.
In contrast, the prediction given by \(\alpha_0\) does not conform to the simulation results.

\begin{figure}
\includegraphics[width=7cm]{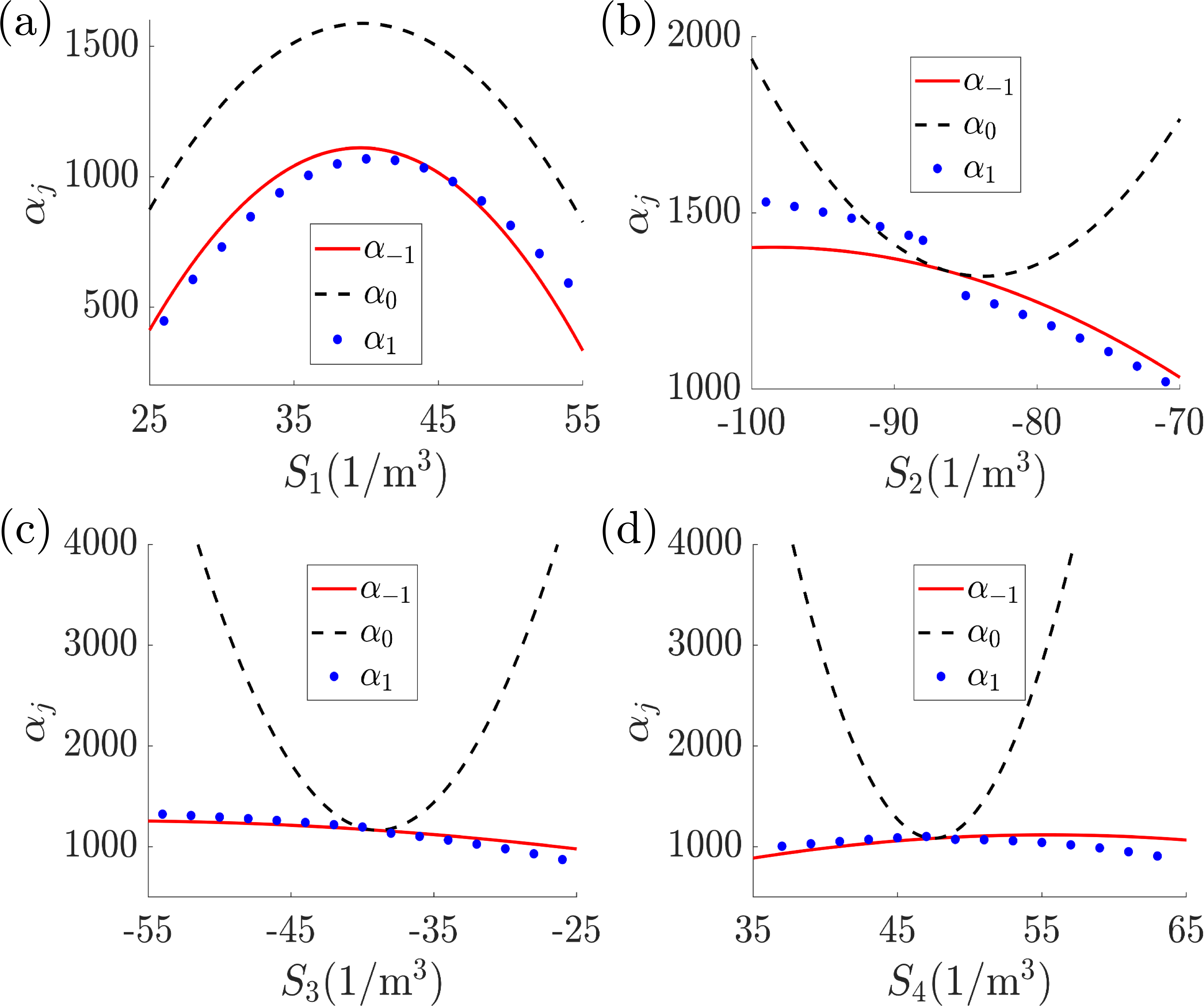}
\caption{\label{fig4}  The empirical nonlinear detuning parameter \(\alpha_{1}\) (blue dots) when (a) \(S_1\) (b) \(S_2\) (c) \(S_3\) (d) \(S_4\) sextupole strength are varied. 
Also shown are \(\alpha_{-1}\) (red lines) and \(\alpha_0\) (black dashed lines). 
}
\end{figure}

To further contrast the analytical fidelity of \(\alpha_{-1}\) to that of \(\alpha_0\), a parametric study was conducted by scanning the strengths of the sextupoles \(S_{1-4}\). Varying these strengths effectively changes the morphology of the Poincare section in Fig. \ref{fig3}. Then, the value of the detuning parameter that renders the contour to align exactly with the simulated Poincare section is dubbed \(\alpha_{1}\), i.e., \(\alpha_{1}\) is the empirical detuning parameter (see details in Appendix C). \(\alpha_{1}\) for different \(S_{1-4}\) are plotted in Fig. \ref{fig4} (blue dots).
Also shown in Fig. \ref{fig4} are \(\alpha_{-1}\) (red lines) and \(\alpha_0\) (black dashed lines) as a function of the sextupole strengths. The dependency of \(\alpha_{-1}\) and \(\alpha_{0}\) on \(S_{1-4}\) can be written as:
\begin{eqnarray}
\alpha_0\left(S_k\right)=a_{2,k}S_k^2+2a_{1,k}S_k+a_{0,k},\\
\alpha_{-1}\left(S_k\right)=b_{2,k}S_k^2+2b_{1,k}S_k+b_{0,k},   
\end{eqnarray}
where \(k\in\{1,2,3,4\}\). These coefficients are derived in Appendix D. 
For instance in Fig. \ref{fig4}(c), the coefficients are calculated as \(a_{2,3}=17.887\), \(b_{2,3}=-0.251\). It is clear that \(\alpha_{-1}\) agrees with \(\alpha_{1}\) better than \(\alpha_0\) does.

\begin{figure}
\includegraphics[width=7cm]{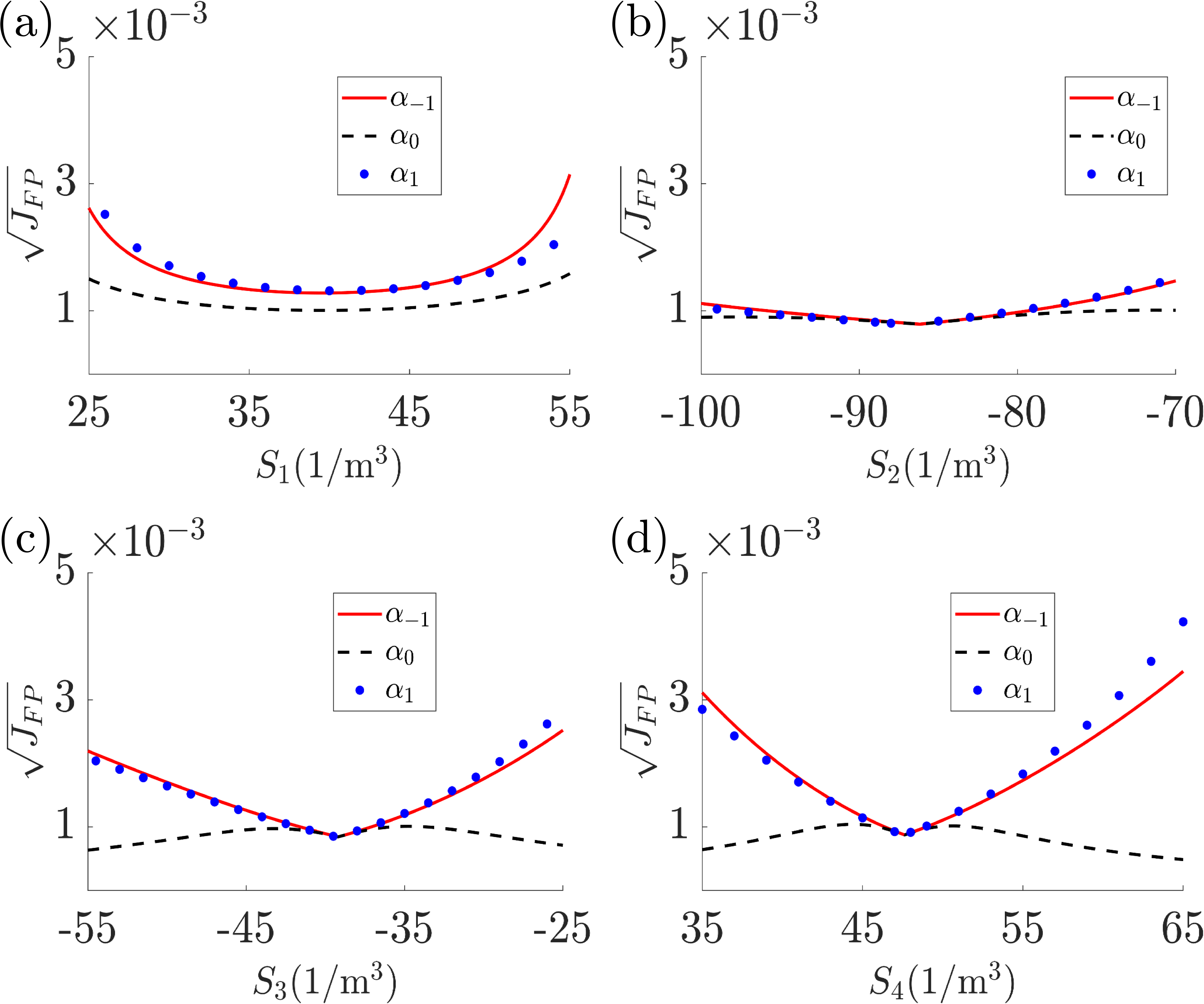}
\caption{\label{fig5}  Fixed points (blue dots) for varying (a) \(S_1\) (b) \(S_2\) (c) \(S_3\) (d) \(S_4\). 
The theoretical prediction of Eq. (\ref{doc17}) with \(\alpha_{-1}\) is plotted in red and that with \(\alpha_0\) is plotted in black.}
\end{figure}

Another analytical prediction that the detuning parameter gives is the location of the fixed point of the rightmost island. 
The distance from the origin to the fixed point can be analytically derived from Eq. (\ref{doc15}) and is given by \cite{15Lee}, 
\begin{equation}
\label{doc17}
\sqrt{J_{FP}}=\left|\frac{g_{3,0,l_{3\nu_x}}}{\alpha_j}\right|\left(\frac{3}{4}+\frac{3}{4}\sqrt{1-\frac{16\alpha_j\delta_\nu}{9g_{3,0,l_{3\nu_x}}^2}}\right).
\end{equation}
The theoretical predictions of Eq. (\ref{doc17}) with \(\alpha_{-1}\) (red lines) and \(\alpha_0\) (black dashed lines) are plotted in Fig. \ref{fig5}. Again, the better agreement between \(\alpha_{-1}\) and \(\alpha_{1}\) is clear.

In Fig. \ref{fig5}, there are regions in parameter space where \(\alpha_{-1}\) and \(\alpha_0\) yields similar predictions (for example, around \(S_3\)=\(-40\) or \(S_4=48\)). 
This is because these are regions where \(g_{3,0,l_{3\nu_x}}\) changes sign and therefore its magnitude becomes small.
Then, at regions where \(g_{3,0,l_{3\nu_x}}\ll\sqrt{\delta_\nu},\  \alpha_{-1}\simeq\alpha_0\) by Eq. (\ref{doc14}) and so the two detuning parameters are indistinguishable.
This can actually be seen in Fig. \ref{fig4} as well;
the regions in question correspond to regions where \(\alpha_{-1}\simeq\alpha_0\). 

From the results presented in Figs. \ref{fig4} and \ref{fig5}, we conclude that \(\alpha_{-1}\) is a much better predictor of TRIBs in storage rings than \(\alpha_0\). 
It is also valid for a much wider range of sextupole strengths or when \(g_{3,0,l_{3\nu_x}}\) is large.
The simulation results shown so far are based on the lattice of PLS-II. 
However, because the only assumption behind the derivation of \(\alpha_{-1}\) is that the tune is near the third-integer resonance, the theoretical result can be applied to any storage ring lattice. 
For instance, our predictions were also checked favorably against simulation results based on the BESSY-II lattice \cite{22BESSY}, which are not presented here.

There are still some discrepancies between the prediction by \(\alpha_{-1}\) and the simulation results.
These differences may come from higher order terms in the perturbation or from the approximations used in the derivation of the Hamiltonian in Eq. (\ref{doc11}). 
As in the case of Ref. \cite{13Soutome}, the higher order terms can in principle be calculated and is left here for future work. Coupling with other motional degrees of freedom will also be left for future work, although for flat beams the present theory should suffice.
\section{SUMMARY}
In summary, we have derived a revised detuning parameter that is well-behaved near third-integer resonance, in contradistinction to the conventional parameter which diverges near this critical point. The resultant Hamiltonian accurately predicts the morphology of transverse resonance island buckets, which are crucial for advanced storage ring operations. This new theory paves the way for the previously-inaccessible, systematic optimization of island sizes and locations in phase-space and reduces unnecessary efforts in haphazard empirical searches for secondary stable orbits.

\begin{acknowledgements}
The authors would like to give special appreciation to Dr. K. Soutome in Japan Synchrotron Radiation Research Institute for his kind introduction for the Hamiltonian dynamics.
The authors would like to appreciate to Dr. J. Y. Lee in PAL for his aid for the simulation in early stage of this study.
The authors would like to extend special thanks to Dr. P. Goslawski, Mr. M. Arlandoo and Dr. J. -G. Hwang in BESSY-II for the many helpful discussions. 
This work was supported National R\&D Program (RS-2022-00154676) and partly by Basic Science Research Program (2021R1F1A105123611) through the National Research Foundation of Korea (NRF) funded by the Ministry of Science and ICT. Y.D.Y. was supported by an appointment to the JRG Program at the APCTP through the Science
and Technology Promotion Fund and Lottery Fund of the Korean Government, and also by Korean
local governments—Gyeongsangbuk-do province and Pohang city. Y.D.Y. was also supported by the
POSCO Science Fellowship of POSCO TJ Park Foundation. 
\end{acknowledgements}
\setcounter{equation}{0}
\setcounter{figure}{0}
\setcounter{table}{0}

\makeatletter


\renewcommand{\theequation}{A\arabic{equation}}
\renewcommand{\thefigure}{A\arabic{figure}}
\setcounter{equation}{0}
\section*{APPENDIX A : Fourier Expansion of Sextupole Potential}

The sextupole potential in the Hamiltonian can be separated into two terms based on the coefficient of \(\psi_2\) in the cosine function, as given by: 
\begin{equation}
V\left(\psi_2,J_2,\theta\right)=RV_1\left(\psi_2,J_2,\theta\right)+RV_2\left(\psi_2,J_2,\theta\right),
\end{equation}
where
\begin{equation}
V_1=\left(\sqrt{J_2}\right)^3\frac{m_x(\theta)}{6\sqrt2}\left(\sqrt{\beta_x(\theta)}\right)^3\cos\left(3\psi_2-3\nu_x\theta+3\chi_x(\theta)\right),
\end{equation}
\begin{equation}
V_2=\left(\sqrt{J_2}\right)^3\frac{m_x(\theta)}{2\sqrt2}\left(\sqrt{\beta_x(\theta)}\right)^3\cos\left(\psi_2-\nu_x\theta+\chi_x(\theta)\right).
\end{equation}
We can separate the first sub-potential \(V_1\) as follows:
\begin{equation}
V_1\equiv\left(\sqrt{J_2}\right)^3M_{3,c}(\theta)\cos{\left(3\psi_2\right)}-\left(\sqrt{J_2}\right)^3M_{3,s}(\theta)\sin{\left(3\psi_2\right)},
\end{equation}
where
\begin{equation}
\label{M3c}
M_{3,c}(\theta)\equiv\frac{m_x(\theta)}{6\sqrt2}\left(\sqrt{\beta_x(\theta)}\right)^3\cos{\left(-3\nu_x\theta+3\chi_x(\theta)\right)},
\end{equation}
\begin{equation}
\label{M3s}
M_{3,s}(\theta)\equiv\frac{m_x(\theta)}{6\sqrt2}\left(\sqrt{\beta_x(\theta)}\right)^3\sin{\left(-3\nu_x\theta+3\chi_x(\theta)\right)}.
\end{equation}
We can express the second sub-potential \(V_2\) in a similar manner as follows:
\begin{equation}
V_2\equiv\left(\sqrt{J_2}\right)^3M_{1,c}(\theta)\cos{\left(\psi_2\right)}-\left(\sqrt{J_2}\right)^3M_{1,s}(\theta)\sin{\left(\psi_2\right)},
\end{equation}
where
\begin{equation}
\label{M1c}
M_{1,c}(\theta)\equiv\frac{m_x(\theta)}{2\sqrt2}\left(\sqrt{\beta_x(\theta)}\right)^3\cos{\left(-\nu_x\theta+\chi_x(\theta)\right)},
\end{equation}
\begin{equation}
\label{M1s}
M_{1,s}(\theta)\equiv\frac{m_x(\theta)}{2\sqrt2}\left(\sqrt{\beta_x(\theta)}\right)^3\sin{\left(-\nu_x\theta+\chi_x(\theta)\right)}.
\end{equation}
Eqs. (\ref{M3c}-\ref{M3s}) and (\ref{M1c}-\ref{M1s}), which are periodic functions of $\theta$, contain all the $\theta$-dependency of the sub-potentials $V_1$ and $V_2$, 
Note that $V$ itself is not periodic function of \(\theta\).
Expressing the above functions by Fourier harmonics yields the following expression for the Hamiltonian:
\begin{equation}
\begin{array}{cl}
\displaystyle \mathcal{H}_2\left(\psi_2,J_2,\theta\right)&=  \displaystyle \nu_xJ_2+\left(\sqrt{J_2}\right)^3\sum_{n=-\infty}^{\infty}{g_{3,0,n}\cos{\left(3\psi_2-n\theta+\xi_{3,0,n}\right)}}\\
& \displaystyle+\left(\sqrt{J_2}\right)^3\sum_{n=-\infty}^{\infty}{g_{1,0,n}\cos{\left(\psi_2-n\theta+\xi_{1,0,n}\right)}},
\end{array} 
\end{equation}
where
\begin{align}
\displaystyle g_{3,0,n}\cos{\xi_{3,0,n}}&\displaystyle  =\frac{R}{\pi}\int_{0}^{2\pi}{\frac{m_x(\theta)}{12\sqrt2}\left(\sqrt{\beta_x(\theta)}\right)^3\cos{\left(-3\nu_x\theta+3\chi_x(\theta)+n\theta\right)}d\theta}, \label{olb11} \\
\displaystyle g_{3,0,n}\sin{\xi_{3,0,n}}&\displaystyle  =\frac{R}{\pi}\int_{0}^{2\pi}{\frac{m_x(\theta)}{12\sqrt2}\left(\sqrt{\beta_x(\theta)}\right)^3\sin{\left(-3\nu_x\theta+3\chi_x(\theta)+n\theta\right)}d\theta}, \label{olb12} \\
\displaystyle g_{1,0,n}\cos{\xi_{1,0,n}}& \displaystyle =\frac{R}{\pi}\int_{0}^{2\pi}{\frac{m_x(\theta)}{4\sqrt2}\left(\sqrt{\beta_x(\theta)}\right)^3\cos{\left(-\nu_x\theta+\chi_x(\theta)+n\theta\right)}d\theta}, \label{olb13}\\
\displaystyle g_{1,0,n}\sin{\xi_{1,0,n}}& \displaystyle =\frac{R}{\pi}\int_{0}^{2\pi}{\frac{m_x(\theta)}{4\sqrt2}\left(\sqrt{\beta_x(\theta)}\right)^3\sin{\left(-\nu_x\theta+\chi_x(\theta)+n\theta\right)}d\theta}. \label{olb14}
\end{align}
If \(m_x(\theta)\) and \(\beta_{x}(\theta)\) are distributed mirror-symmetrically, the oddness of integrated function implies that Eqs. (\ref{olb12}) and (\ref{olb14}) are equal to zero. This implies that the Fourier expansion of sextupole potential have a phase of either zero or \(\pi\). Another notable feature is that Equations (\ref{olb11}-\ref{olb14}) can be expressed in complex form as follows:
\begin{equation}
\label{olb15}
g_{3,0,n}e^{i\xi_{3,0,n}}=\frac{\sqrt2 R}{24\pi}\int_{0}^{2\pi}{m_x(\theta)\left(\sqrt{\beta_x(\theta)}\right)^3\exp\left(i\left(-(3\nu_x-n)\theta+3\chi_x(\theta)\right)\right)d\theta},
\end{equation}
\begin{equation}
\label{olb16}
g_{1,0,n}e^{i\xi_{1,0,n}}=\frac{\sqrt2 R}{8\pi}\int_{0}^{2\pi}{m_x(\theta)\left(\sqrt{\beta_x(\theta)}\right)^3\exp\left(i\left(-\left(\nu_x-n\right)\theta+\chi_x(\theta)\right)\right)d\theta}.
\end{equation}

\renewcommand{\theequation}{B\arabic{equation}}
\renewcommand{\thefigure}{B\arabic{figure}}
\setcounter{equation}{0}
\section*{APPENDIX B : Proof of Equivalence Between Two Detuning Parameters \(\alpha_{x,x}\) and \(\alpha_{0}\)}
In this section, we present a proof for the equivalence between two parameters
\begin{equation}
\label{alpha0}
\alpha_0=\displaystyle\frac{3}{2}\left(\displaystyle\sum_{n=-\infty}^{\infty}{3f_{3,0,n}g_{3,0,n}}+\sum_{n=-\infty}^{\infty}{f_{1,0,n}g_{1,0,n}}\right),
\end{equation}
and
\begin{equation}
\label{alphaxx}
\begin{array}{cl}
\displaystyle \alpha_{x,x} &= \displaystyle-\frac{1}{64\pi}\int_{0}^{L}{ds\ m_x(s)\beta_x^{\frac{3}{2}}(s)}\\
&\displaystyle \times {\int_{s}^{s+L}m_{x}(s^\prime)\beta_x^{\frac{3}{2}}(s^\prime)\left[\frac{\cos{3\Psi_x(s^\prime,s)}}{\sin{3\pi\nu_x}}+\frac{3\cos{\Psi_x(s^\prime,s)}}{\sin{\pi\nu_x}}\right]ds^\prime},
\end{array}
\end{equation}
where
\begin{equation}
\displaystyle \Psi_x(s^\prime,s)=\chi_x(s^\prime)-\chi_x(s)-\pi\nu_x.
\end{equation}
Eq. (\ref{alpha0}) is derived in main article, Eq. (\ref{alphaxx}) is the well-known nonlinear detuning parameter. By applying the delta function approximation for the sextupole strength \(m_{x}(s)\) into Eq. ({\ref{alphaxx}}), we show that the above equation is equivalent to Eq. (196) in \cite{15Lee}. 
To prove the equivalence, we separate integral form of the nonlinear detuning parameter \(\alpha_{x,x}\) into two terms as follows:
\begin{align}
\displaystyle \alpha_{x,x,3} & \displaystyle =  -\frac{1}{64\pi}\int_{0}^{L}{ds\ m_x(s)\beta_x^{\frac{3}{2}}(s)\cdot\int_{s}^{s+L}m_x(s^\prime)\beta_x^{\frac{3}{2}}(s^\prime)\frac{\cos{3\Psi_x(s^\prime,s)}}{\sin{3\pi\nu_x}}\ ds^\prime},\\
\displaystyle \alpha_{x,x,1} & \displaystyle =-\frac{3}{64\pi}\int_{0}^{L}{ds\ m_x(s)\beta_x^{\frac{3}{2}}(s)\cdot\int_{s}^{s+L}m_x(s^\prime)\beta_x^{\frac{3}{2}}(s^\prime)\frac{\cos{\Psi_x(s^\prime,s)}}{\sin{\pi\nu_x}}\ ds^\prime}.
\end{align}
We express \(\alpha_{x,x,3}\) as follows to facilitate further calculation,
\begin{equation}
\label{alpha03}
\begin{array}{cl}
\alpha_{x,x,3}&=\displaystyle-\frac{1}{128\pi}\left(\int_{0}^{L}{ds\ m_x(s)\beta_x^{\frac{3}{2}}(s)e^{-i3\chi_x(s)}\cdot\int_{s}^{s+L}m_{x}(s^\prime)\beta_x^{\frac{3}{2}}(s^\prime)e^{-i 3\nu_x\left(\frac{s^\prime-s}{R}\right)}\frac{e^{i\mathrm{\Upsilon}_x\left(s,s^\prime\right)}}{\sin{3\pi\nu_x}}ds^\prime\ }\right.\\
&\displaystyle\left.+\int_{0}^{L}{ds\ m_x(s)\beta_x^{\frac{3}{2}}(s)e^{i3\chi_x(s)}\cdot\int_{s}^{s+L}m_{x}(s^\prime)\beta_x^{\frac{3}{2}}(s^\prime)e^{i\cdot3\nu_x\left(\frac{s^\prime-s}{R}\right)}\frac{e^{-i\mathrm{\Upsilon}_x\left(s,s^\prime\right)}}{\sin{3\pi\nu_x}}ds^\prime}\right),
\end{array}    
\end{equation}
where
\begin{equation}
\label{docd5}
\mathrm{\Upsilon}_x\left(s,s^\prime\right)\equiv3\chi_x(s^\prime)-3\pi\nu_x+3\nu_x\left(\frac{s^\prime-s}{R}\right).
\end{equation}
If we apply following relation \cite{16Merminga}
\begin{equation}
\label{docd6}
\sum_{n=-\infty}^{\infty}\frac{e^{i\left(n\theta+b\right)}}{n-3\nu_x}=-\frac{\pi}{\sin{3\pi\nu_x}}e^{i\left(b+3\nu_x\left(\theta-\pi\right)\right)} ,
\end{equation}
where $3\nu_x$ is not an integer, we can express the Eq. (\ref{alpha03}) as follows:
\begin{equation}
\begin{array}{cl}
\alpha_{x,x,3} = &\displaystyle\frac{1}{128\pi^2}\displaystyle\sum_{n=-\infty}^{\infty}\left\{\int_{0}^{L}{ds\ m_x(s)\beta_x^{\frac{3}{2}}(s)e^{i\left(-3\chi_x(s)+\frac{3\nu_xs}{R}-\frac{ns}{R}\right)}} \right.\\ 
&\displaystyle \times \int_{s}^{s+L}{m_{x}(s^\prime)\beta_x^{\frac{3}{2}}(s^\prime)\frac{ e^{i\left(\frac{n s^\prime}{R}-\frac{3\nu_x s^\prime}{R}+3\chi_x(s^\prime)\right)} }{n-3\nu_x}ds^\prime}+\int_{0}^{L}{ds\ m_x(s)\beta_x^{\frac{3}{2}}(s)e^{i\left(3\chi_x(s)-\frac{3\nu_xs}{R}+\frac{ns}{R}\right)}} \\
&\displaystyle \left. \times \int_{s}^{s+L}m_{x}(s^\prime)\beta_x^{\frac{3}{2}}(s^\prime)\frac{e^{-i\left(\frac{n s^\prime}{R}-\frac{3\nu_x s^\prime}{R}+3\chi_x(s^\prime)\right)}}{n-3\nu_x}ds^\prime\right\}.
\end{array}    
\end{equation}
The exponential function can be modified as follows due to the periodicity of the internal functions:
\begin{equation}
\displaystyle
e^{i\left(n\frac{s^\prime+L}{R}-3\nu_x\left(\frac{s^\prime+L}{R}\right)+3\chi_x\left(s^\prime+L\right)\right)}=e^{i\left(n\frac{s^\prime}{R}-3\nu_x\left(\frac{s^\prime}{R}\right)+3\chi_x(s^\prime)\right)}.
\end{equation}
Using the relation in Eq. (\ref{docd6}), we can express Eq. (\ref{alpha03}) by using Eqs. (\ref{olb15}) and (\ref{olb16}) as follows: 
\begin{equation}
\begin{array}{cl}
\label{old10}
\alpha_{x,x,3}&=\displaystyle\frac{1}{128\pi^2}\displaystyle\sum_{n=-\infty}^{\infty}\left\{\int_{0}^{L}{ds\ m_x(s)\beta_x^{\frac{3}{2}}(s)e^{i\left(-3\chi_x(s)+\frac{3\nu_xs}{R}-\frac{ns}{R}\right)}} \right.\\ 
&\displaystyle ~ ~ \times \int_{0}^{L}{m_{x}(s^\prime)\beta_x^{\frac{3}{2}}(s^\prime)\frac{e^{-i\left(-3\chi_x(s^\prime)+\frac{3\nu_x s^\prime}{R}-\frac{n s^\prime}{R}\right)}}{n-3\nu_x}ds^\prime}+\int_{0}^{L}{ds\ m_x(s)\beta_x^{\frac{3}{2}}(s)e^{i\left(3\chi_x(s)-\frac{3\nu_xs}{R}+\frac{ns}{R}\right)}} \\
&\displaystyle ~ ~ \left. \times \int_{0}^{L}m_{x}(s^\prime)\beta_x^{\frac{3}{2}}(s^\prime)\frac{e^{i\left(-3\chi_x(s^\prime)+\frac{3\nu_x s^\prime}{R}-\frac{n s^\prime}{R}\right)}}{n-3\nu_x}ds^\prime\right\} \\
&\displaystyle=\frac{3}{2}\sum_{n=\infty}^{\infty}\left(3g_{3,0,n}\cdot f_{3,0,n}\right).
\end{array}    
\end{equation}
where \(f_{3,0,n}\) is given in Eq. (\ref{olc16}).
By following a similar calculation process, we can also obtain the following relation,
\begin{equation}
\displaystyle\alpha_{x,x,1}=\frac{3}{2}\sum_{n=-\infty}^{\infty}\left(g_{1,0,n}\cdot f_{1,0,n}\right).
\end{equation}
Thus, the nonlinear detuning parameter \(\alpha_{x,x}\) is expressed by
\begin{equation}
\displaystyle\alpha_{x,x}=\alpha_{x,x,3}+\alpha_{x,x,1}=\frac{3}{2}\sum_{n=\infty}^{\infty}\left(3g_{3,0,n}\cdot f_{3,0,n}+g_{1,0,n}\cdot f_{1,0,n}\right)=\alpha_{0}.
\end{equation}
As a result, we have demonstrated the equivalence of the two nonlinear detuning parameters given in Eqs. (\ref{alpha0}) and (\ref{alphaxx}).
\renewcommand{\theequation}{C\arabic{equation}}
\renewcommand{\thefigure}{C\arabic{figure}}
\setcounter{equation}{0}
\setcounter{figure}{0}

\section*{APPENDIX C : Determination of the Empirical Detuning Parameter \( \alpha_{1} \) }

\begin{figure}
\includegraphics[width=7cm]{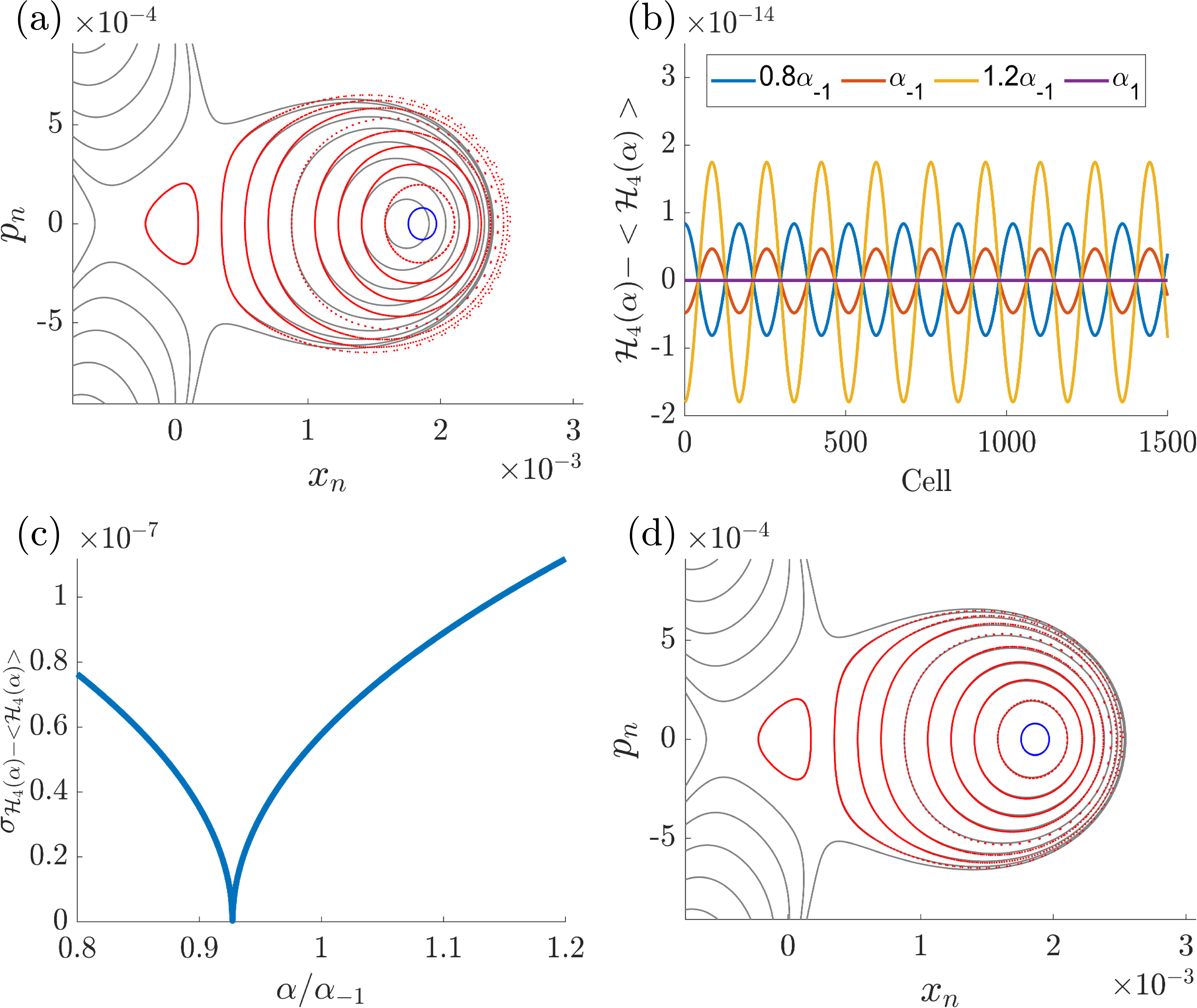}
\caption{\label{fige1} \(\alpha_{1}\) calculating process 
(a) Tracking results of electron for 1500 cells and contour plot. 
The tracking was performed for every 3 cells to confine the results to a single island.
The blue line represents the tracking results near the center of the island buckets. 
(b) Oscillation part of \(\mathcal{H}_4\) for blue line of Fig. \ref{fige1}(a). It is defined by \(\left(\mathcal{H}_4-<\mathcal{H}_4>\ \right)\)
(c) We calculated the standard deviation of the oscillation part of \(\mathcal{H}_4\) for the blue line in Fig. \ref{fige1}(a) and defined the value of \(\alpha\) that minimizes the function \(\sigma_{\mathcal{H}_4 - \left<\mathcal{H}_4\right>}\) as \(\alpha_{1}\). \(\alpha_{-1}\) in \(x\)-axis is 1058.1 and \(\alpha_{1}\) is 981.9. 
(d) Graph of tracking results and Hamiltonian contours in (a) and plot of Hamiltonian defined by using \(\alpha_{1}\).}
\end{figure}

This section describes the process used to determine the value of \(\alpha_{1}\) in this study. 
First, a tracking simulation was performed using an arbitrary value on the \(x\)-axis as the initial point.
This simulation was conducted over 1500 cells, using the lattice presented in Fig. \ref{fig2} of main article. 
To ensure consistency, results were stored every 3 cells, ensuring all tracking simulation results were in the same island bucket. 
The results are shown in Fig. \ref{fige1}(a).

Then, the point on the \(x\)-axis closest to the fixed point of the island buckets was identified from the tracking results. 
The blue line in Fig. \ref{fige1}(a) depicts the tracking result of the electron closest to the fixed point.
Following this, the Hamiltonian was redefined as a function of \(\alpha\) and the number of cells which electron passed and it is given by
\begin{equation}
\displaystyle
\mathcal{H}_4=\delta J+g_{3,0,l_{3\nu_x}}J^\frac{3}{2}\cos{\left(3\psi+\xi_{3,0,l_{3\nu_x}}\right)}+\frac{1}{2}\alpha J^2
\end{equation}
The graph of the Hamiltonian's oscillation part for the electron closest to the fixed point at each \(\alpha\) was obtained. 
This oscillation part was calculated by subtracting the mean value of the Hamiltonian with respect to its position from the Hamiltonian itself. 
The graph of the oscillation part of the Hamiltonian is shown in Fig. \ref{fige1}(b).

The value of \(\alpha_{1}\) was defined by computing the standard deviation of the oscillation part of the \(\mathcal{H}_4\) for each \(\alpha\) value and selecting the point where \(\sigma_{\mathcal{H}_4-<\mathcal{H}_4>}\) is minimized as \(\alpha_{1}\).
This selection was made because the same particle has the same \(\mathcal{H}_4\) values. 
Fig. \ref{fige1}(c) shows the graph of the standard deviation of the oscillation part.
The \(\alpha\) value which has the minimum standard deviation corresponds to the value of \(\alpha_{1}\).
Finally, \(\alpha_{1}\) was used to redraw the contour of the tracking results, and the Hamiltonian plot with \(\alpha_{1}\) is depicted in Fig. \ref{fige1}(d).
This plot demonstrates that all electrons on the contour can be covered. Therefore, we can assume that \(\alpha_{1}\)is a reliable value.

\renewcommand{\theequation}{D\arabic{equation}}
\renewcommand{\thefigure}{D\arabic{figure}}
\setcounter{equation}{0}
\section*{APPENDIX D : Calculation of the Nonlinear Detuning Parameters as a Function of Sextupole Strength in a Mirror-Symmetric Lattice}
This section presents the derivation of the detuning parameters \(\alpha_{-1}\) and \(\alpha_0\) as functions of sextupole strength, for a symmetric ring. Equation (\ref{alphaxx}) can be simplified by removing the s-dependence of the integrand over the \({ds}^\prime\) integration range.
\begin{equation}
\begin{array}{cl}
\displaystyle
\alpha_0= & \displaystyle -\frac{1}{64\pi}\int_{0}^{L}{\int_{0}^{L}{m_{x}(s^\prime)m_x(s)\beta_x^{\frac{3}{2}}(s^\prime)\beta_x^{\frac{3}{2}}(s) }} \\ 
&\displaystyle \times \left(\frac{\cos{3\left(\pi\nu_x-\left|\chi_x(s^\prime)-\chi_x(s)\right|\right)}}{\sin{3\pi\nu_x}}+\frac{3\cos{\left(\pi\nu_x-\left|\chi_x(s^\prime)-\chi_x(s)\right|\right)}}{\sin{\pi\nu_x}}\right)ds^\prime d s .
\end{array}    
\end{equation}
To exploit the lattice symmetry, the integral range in the above equation can be shifted by \(\frac{L}{2}\), resulting in the following expression:
\begin{equation}
\begin{array}{cl}
\label{docF1}
\displaystyle
\alpha_0= & \displaystyle -\frac{1}{64\pi}\int_{-\frac{L}{2}}^{\frac{L}{2}}{\int_{-\frac{L}{2}}^{\frac{L}{2}}{m_{1}(s^\prime)m_{1}(s)\beta_{x,1}^{\frac{3}{2}}(s^\prime)\beta_{x,1}^{\frac{3}{2}}(s)  }}\\ 
&\displaystyle \times \left(\frac{\cos{3\left(\pi\nu_x-\left|\chi_x(s^\prime)-\chi_x(s)\right|\right)}}{\sin{3\pi\nu_x}}+\frac{3\cos{\left(\pi\nu_x-\left|\chi_x(s^\prime)-\chi_x(s)\right|\right)}}{\sin{\pi\nu_x}}\right)ds^\prime d s .
\end{array}    
\end{equation}
where \(m_1(s)=m_{x}\left(s+\frac{L}{2}\right)\), \(\beta_{x,1}(s)=\beta_x\left(s+\frac{L}{2}\right)\).
If the integral expression is rearranged such that the integral interval is limited to 0 to \(\frac{L}{2}\), the resulting equation is as follows:
\begin{equation}
\begin{array}{cl}
\displaystyle
\alpha_0&\displaystyle=-\displaystyle\frac{2}{64\pi}\int_{0}^{\frac{L}{2}}{\int_{0}^{\frac{L}{2}}{m_1(s^\prime)m_1(s)\beta_{x,1}^{\frac{3}{2}}(s^\prime)\beta_{x,1}^{\frac{3}{2}}(s)}} \\
& \displaystyle \left[\left\{\cot{3\pi\nu_x}\left(\cos{3\left(\chi_x(s^\prime)-\chi_x(s)\right)}+\cos{3\left(\chi_x(s^\prime)+\chi_x(s)\right)}\right)+\sin{3\left|\chi_x(s^\prime)-\chi_x(s)\right|} \right.\right. \\
&\left.\left.+\sin{3\left(\chi_x(s^\prime)+\chi_x(s)\right)}\right\}+3\left\{\cot{\pi\nu_x}\left(\cos{\left(\chi_x(s^\prime)-\chi_x(s)\right)}+\cos{\left(\chi_x(s^\prime)+\chi_x(s)\right)}\right) \right.\right.\\
&\displaystyle \left.\left.+\sin{\left|\chi_x(s^\prime)-\chi_x(s)\right|}+\sin{\left(\chi_x(s^\prime)+\chi_x(s)\right)}\right\}\right]ds^\prime ds,
\end{array}    
\end{equation}
where \(s_i\) is a positive position of \(i\)-th sextupole magnet, \(l_i\) is a length of \(i\)-th sextupole magnet, \(S_i\) is a strength of sextupole magnet and \(N\) is the number of sextupole magnets pairs.
The index is arranged according to the distance from the origin and is restricted to sextupole magnets situated in the positive position due to their symmetrical distribution.
Expressing the above equation as a quadratic function for the \(k\)-th sextupole strength yields the following result:
\begin{equation}
\displaystyle
\alpha_0\left(S_k\right)=a_{2,k}S_k^2+2a_{1,k}S_k+a_{0,k},
\end{equation}
where \(a_{2,k}\) is defined by
\begin{equation}
\begin{array}{cl}
\displaystyle a_{2,k}=&  \displaystyle -\frac{2}{64\pi}\int_{s_k}^{s_k+l_k}{\int_{s_k}^{s_k+l_k}{ \beta_{x,1}^{\frac{3}{2}}(s^\prime) \beta_{x,1}^{\frac{3}{2}}(s) } }\\
&\displaystyle \left[  2\cot{3\pi\delta} \cos{3\chi_x(s)}\cos{3\chi_x(s^\prime)}+\sin{3|\chi_x(s^\prime)-\chi_x(s)|}+\sin{3(\chi_x(s^\prime)+\chi_x(s)|)}   \right. \\ 
&\displaystyle \left. +3\left( 2\cot{\pi\nu_x}\cos{\chi_x(s^\prime)}\cos{\chi_x(s)}+\sin{|\chi_x(s^\prime)-\chi_x(s)|}+\sin{(\chi_x(s^\prime)+\chi_x(s))} \right) \right]ds^\prime ds,
\end{array}    
\end{equation}
\(a_{1,k}\) is defined by 
\begin{equation}
\begin{array}{cl}
\displaystyle a_{1,k}=&  \displaystyle -\frac{2}{64\pi}\sum_{i=1,i\neq k}^{N}{S_i \int_{s_k}^{s_k+l_k}{\int_{s_i}^{s_i+l_i}{ \beta_{x,1}^{\frac{3}{2}}(s^\prime) \beta_{x,1}^{\frac{3}{2}}(s) } }}\\
&\displaystyle \left[  2\cot{3\pi\delta} \cos{3\chi_x(s)}\cos{3\chi_x(s^\prime)}+\sin{3|\chi_x(s^\prime)-\chi_x(s)|}+\sin{3(\chi_x(s^\prime)+\chi_x(s)|)}   \right. \\ 
&\displaystyle \left. +3\left( 2\cot{\pi\nu_x}\cos{\chi_x(s^\prime)}\cos{\chi_x(s)}+\sin{|\chi_x(s^\prime)-\chi_x(s)|}+\sin{(\chi_x(s^\prime)+\chi_x(s))} \right) \right]ds^\prime ds,
\end{array}    
\end{equation}
and \(a_{0,k}\) is defined by 
\begin{equation}
\begin{array}{cl}
\displaystyle a_{0,k}=&  \displaystyle -\frac{2}{64\pi}\sum_{i,j=1,i,j\neq k}^{N}{S_i S_j \int_{s_j}^{s_j+l_j}{\int_{s_i}^{s_i+l_i}{ \beta_{x,1}^{\frac{3}{2}}(s^\prime) \beta_{x,1}^{\frac{3}{2}}(s) } }}\\
&\displaystyle \left[  2\cot{3\pi\delta} \cos{3\chi_x(s)}\cos{3\chi_x(s^\prime)}+\sin{3|\chi_x(s^\prime)-\chi_x(s)|}+\sin{3(\chi_x(s^\prime)+\chi_x(s)|)}   \right. \\ 
&\displaystyle \left. +3\left( 2\cot{\pi\nu_x}\cos{\chi_x(s^\prime)}\cos{\chi_x(s)}+\sin{|\chi_x(s^\prime)-\chi_x(s)|}+\sin{(\chi_x(s^\prime)+\chi_x(s))} \right) \right]ds^\prime ds.
\end{array}    
\end{equation}
For a symmetric cell, the expression for \(g_{3,0,l_{3\nu_x}}\) is given by
\begin{equation}
\displaystyle
g_{3,0,{l_{3\nu}}_x}=\frac{\sqrt2}{12\pi}\sum_{i=1}^{N}{S_i\oint{\beta_{x,1}^{\frac{3}{2}}(s)\cos{\left(3\chi_x(s)-3\left(\nu_x-\frac{l_{3\nu_x}}{3}\right)\frac{s}{R}\right)}ds}}.
\end{equation}
Hence, the expression for \(g_{3,0,l_{3\nu_x}}^2\) is given by
\begin{equation}
\begin{array}{cl}
\displaystyle
\frac{3}{2\delta}g_{3,0,{l_{3\nu}}_x}^2=&\displaystyle\frac{2}{96\delta_\nu\pi^2}\sum_{i,j=1}^{N}{S_iS_j\int_{s_j}^{s_j+l_j}\int_{s_i}^{s_i+l_i}}\\
&\displaystyle \times {{\beta_{x,1}^{\frac{3}{2}}(s^\prime)\beta_{x,1}^{\frac{3}{2}}(s)\cos{\left(3\chi_x(s^\prime)-3\delta\frac{s^\prime}{R}\right)}\cos{\left(3\chi_x(s)-3\delta\frac{s}{R}\right)}ds^\prime d s}}.
\end{array}  
\end{equation}
Thus, the value of \(\alpha_{-1}\) is obtained as follow:
\begin{equation}
\begin{array}{cl}
\displaystyle\alpha_{-1} = &\displaystyle-\frac{1}{32\pi}\sum_{i,j=1}^{N}{S_iS_j\int_{s_j}^{s_j+l_j}\int_{s_i}^{s_i+l_i}{\beta_{x,1}^{\frac{3}{2}}(s^\prime)\beta_{x,1}^{\frac{3}{2}}(s)\left(2\cot{3\pi\nu_x}\cos{3\chi_x(s)}\cos{3\chi_x(s^\prime)}\right.}}\\
     &\displaystyle-\frac{2}{3\delta\pi}\cos{\left(3\chi_x(s^\prime)-3\delta\frac{s^\prime}{R}\right)}\cos{\left(3\chi_x(s)-3\delta\frac{s}{R}\right)}+\sin{3\left|\chi_x(s^\prime)-\chi_x(s)\right|}\\
     &\left.+\sin{3\left(\chi_x(s^\prime)+\chi_x(s)\right)}+3\left(2\cot{\pi\nu_x}\cos{\chi_x(s)}\cos{\chi_x(s^\prime)}+\sin{\left|\chi_x(s^\prime)-\chi_x(s)\right|} \right.\right.\\ 
     &\left.\left.+\sin{\left(\chi_x(s^\prime)+\chi_x(s)\right)}\right)\right)ds^\prime ds.
\end{array}    
\end{equation}
Expressing the above equation as a quadratic function for the \(k\)-th sextupole strength yields the following result:
\begin{equation}
\displaystyle \alpha_{-1}\left(S_k\right)=b_{2,k}S_k^2+2b_{1,k}S_k+b_{0,k},
\end{equation}
where \(b_{1,k}\) is defined by 
\begin{equation}
\begin{array}{cl}
\displaystyle b_{1,k} = &\displaystyle-\frac{1}{32\pi}\sum_{i=1,i\neq k}^{N}{S_i\int_{s_i}^{s_i+l_i}\int_{s_k}^{s_k+l_k}{\beta_{x,1}^{\frac{3}{2}}(s^\prime)\beta_{x,1}^{\frac{3}{2}}(s)\left(2\cot{3\pi\nu_x}\cos{3\chi_x(s)}\cos{3\chi_x(s^\prime)}\right.}}\\
     &\displaystyle-\frac{2}{3\delta\pi}\cos{\left(3\chi_x(s^\prime)-3\delta\frac{s^\prime}{R}\right)}\cos{\left(3\chi_x(s)-3\delta\frac{s}{R}\right)}+\sin{3\left|\chi_x(s^\prime)-\chi_x(s)\right|}\\
     &\left.+\sin{3\left(\chi_x(s^\prime)+\chi_x(s)\right)}+3\left(2\cot{\pi\nu_x}\cos{\chi_x(s)}\cos{\chi_x(s^\prime)}+\sin{\left|\chi_x(s^\prime)-\chi_x(s)\right|} \right.\right.\\ 
     &\left.\left.+\sin{\left(\chi_x(s^\prime)+\chi_x(s)\right)}\right)\right)ds^\prime ds,
\end{array}    
\end{equation}
\(b_{0,k}\) is defined by 
\begin{equation}
\begin{array}{cl}
\displaystyle b_{0,k} = &\displaystyle-\frac{1}{32\pi}\sum_{i,j=1,i,j\neq k}^{N}{S_i S_j \int_{s_i}^{s_i+l_i}\int_{s_j}^{s_j+l_j}{\beta_{x,1}^{\frac{3}{2}}(s^\prime)\beta_{x,1}^{\frac{3}{2}}(s)\left(2\cot{3\pi\nu_x}\cos{3\chi_x(s)}\cos{3\chi_x(s^\prime)}\right.}}\\
     &\displaystyle-\frac{2}{3\delta\pi}\cos{\left(3\chi_x(s^\prime)-3\delta\frac{s^\prime}{R}\right)}\cos{\left(3\chi_x(s)-3\delta\frac{s}{R}\right)}+\sin{3\left|\chi_x(s^\prime)-\chi_x(s)\right|}\\
     &\left.+\sin{3\left(\chi_x(s^\prime)+\chi_x(s)\right)}+3\left(2\cot{\pi\nu_x}\cos{\chi_x(s)}\cos{\chi_x(s^\prime)}+\sin{\left|\chi_x(s^\prime)-\chi_x(s)\right|} \right.\right.\\ 
     &\left.\left.+\sin{\left(\chi_x(s^\prime)+\chi_x(s)\right)}\right)\right)ds^\prime ds,
\end{array}    
\end{equation}
and \(b_{2,k}\) is defined by
\begin{equation}
\begin{array}{cl}
\displaystyle b_{2,k} = &\displaystyle -\frac{1}{32\pi}\int_{s_k}^{s_k+l_k}{\int_{s_k}^{s_k+l_k}{ \beta_{x,1}^{\frac{3}{2}}(s^\prime) \beta_{x,1}^{\frac{3}{2}}(s) }} \\
&\displaystyle \times \biggr(\frac{2}{3\pi \delta} \cos{3\chi_x(s) } \cos{3\chi_x(s^\prime) }-\frac{2}{3\pi \delta} \cos{\left(3\chi_x(s)-3\delta \frac{s}{R}\right) } \cos{\left(3\chi_x(s^\prime)-3\delta \frac{s^\prime}{R}\right) } \\
&\displaystyle -2\cos{3\chi_x(s)}\cos{\chi_x(s^\prime)}\pi\delta+\sin{3|\chi_x(s^\prime)-\chi_x(s)|}+\sin{3\left(\chi_x(s^\prime)+\chi_x(s)\right)}\\
&\displaystyle +3\left( 2\cot{\pi\nu_x}\cos{\chi_x(s)}\cos{\chi_x(s^\prime)}+\sin{|\chi_x(s^\prime)-\chi_x(s)|}+\sin{\left(\chi_x(s^\prime)+\chi_x(s)\right)} \right) \biggr).
\end{array}    
\end{equation}
Taking the limit of \(\delta\ \rightarrow\ 0\), we can use next relation,
\begin{equation}
\displaystyle \lim_{\delta\rightarrow0}{\left(\frac{\sin{\left(\frac{3\delta}{2}\frac{s}{R}\right)}}{3\pi\delta}\right)}=\frac{s}{2\pi R}\ .    
\end{equation}
Finally, \(b_{2,k}\) is calculated as, 
\begin{equation}
\begin{array}{cl}
\displaystyle b_{2,k} = &\displaystyle -\frac{1}{32\pi}\int_{s_k}^{s_k+l_k}{\int_{s_k}^{s_k+l_k}{ \beta_{x,1}^{\frac{3}{2}}(s^\prime) \beta_{x,1}^{\frac{3}{2}}(s) }} \\
&\displaystyle \times \biggr(-8 \sin{\left(3\chi_x(s^\prime)-\frac{3\delta}{2}\frac{s^\prime}{R} \right)}\cos{\left(3\chi_x(s)-\frac{3\delta}{2}\frac{s}{R} \right)}\cos{\left(\frac{3\delta}{2}\frac{s}{R} \right)}\frac{s^\prime}{2\pi R} \\
&\displaystyle -2\cos{3\chi_x(s)}\cos{3\chi_x(s^\prime)}\pi\delta+\sin{3|\chi_x(s^\prime)-\chi_x(s)|}+\sin{3\left(\chi_x(s^\prime)+\chi_x(s)\right)}\\
&\displaystyle +3\left( 2\cot{\pi\nu_x}\cos{\chi_x(s)}\cos{\chi_x(s^\prime)}+\sin{|\chi_x(s^\prime)-\chi_x(s)|}+\sin{\left(\chi_x(s^\prime)+\chi_x(s)\right)} \right) \biggr).
\end{array}    
\end{equation}
Due to the absence of \(\delta\) in the denominator, the coefficient \(b_{2,k}\) remains finite for all values of \(\delta\). 

\providecommand{\noopsort}[1]{}\providecommand{\singleletter}[1]{#1}%

\end{document}